\documentclass[twocolumn]{aastex631}
\usepackage{CJKutf8}

\usepackage{lipsum} 
\usepackage{longtable}  
\usepackage{booktabs} 
\usepackage{array} 
\usepackage{amsmath}
\usepackage{appendix}
\usepackage{graphicx}                                                           
\usepackage{float} 

\begin{document}

\title{How Long Will the Quasar UV/Optical Flickering Be Damped? 
\uppercase\expandafter{\romannumeral2}. the Observational Test}

\correspondingauthor{Mouyuan Sun}
\email{msun88@xmu.edu.cn}

\author[0000-0002-1497-8371]{Guowei Ren}
\affiliation{Department of Astronomy, Xiamen University, Xiamen, 
Fujian 361005, China; msun88@xmu.edu.cn}

\author[0009-0005-2801-6594]{Shuying Zhou}
\affiliation{Department of Astronomy, Xiamen University, Xiamen, 
Fujian 361005, China; msun88@xmu.edu.cn}

\author[0000-0002-0771-2153]{Mouyuan Sun}
\affiliation{Department of Astronomy, Xiamen University, Xiamen, 
Fujian 361005, China; msun88@xmu.edu.cn}

\author[0000-0002-1935-8104]{Yongquan Xue}
\affiliation{Department of Astronomy, University of Science and Technology of China, Hefei 230026, China}
\affiliation{School of Astronomy and Space Science, University of Science and Technology of China, Hefei 230026, China}



\begin{abstract}
The characteristic timescale at which the variability of active galactic nuclei (AGNs) turns from red noise to white noise can probe the accretion physics around supermassive black holes (SMBHs). A number of works have studied the characteristic timescale of quasars and obtained quite different scaling relations between the timescale and quasar physical properties. One possible reason for the discrepancies is that the characteristic timescale can be easily underestimated if the light curves are not long enough. In this work, we construct well-defined AGN samples to observationally test the relationships between the characteristic timescale and AGN properties obtained by previous works. Our samples eliminate the effects of insufficient light-curve lengths. We confirm that the timescale predictions \citep{Zhou2024} of the Corona Heated Accretion disk Reprocessing model are consistent with our timescale measurements. The timescale predictions by empirically relations \citep[e.g.,][]{Kelly2009} are systematically smaller than our measured ones. Our results provide further evidence that AGN variability is driven by thermal fluctuations in SMBH accretion disks. Future flagship time-domain surveys can critically test our conclusions and reveal the physical nature of AGN variability.

\end{abstract}

\keywords{Accretion (14); Light curves (918); Quasars (1319); Supermassive black holes (1663)}


\section{Introduction} \label{sec:intro}
Active Galactic Nuclei (AGNs) are the brightest persistent sources in the Universe. They emit non-stellar continuum emission in a broad range of electromagnetic spectrum from radio to $\mathrm{X}$-ray and $\gamma$-ray \citep[for related reviews, see, e.g.,][]{Urry1995, Ulrich1997}. The strong continuum emission, which varies significantly on timescales of minutes, hours, days, and months, is usually believed to originate in the central supermassive black hole (SMBH) accretion disk. The central engine of an AGN is often too compact to resolve directly because of its small physical scale and high redshift. The multi-band inherent and complex variations (i.e., light curves) play an essential role in probing the properties of the central engine, and the physical processes of AGNs \citep[e.g.,][]{Cackett2021}. 

The study of AGN multi-band light curves provides valuable information about the AGN structure and physical properties. By measuring the time lags between light curves at different wavelengths, the sizes of the accretion disk, broad line region, and dusty torus, and even the black hole mass of the SMBH can be measured using the reverberation mapping (RM) technique \citep[e.g.,][]{Blandford1982, Peterson1993, Cackett2021}. The black hole masses of more than 100 AGNs have been measured by the RM technique in many previous works \citep[e.g.,][]{Kaspi2000, Peterson2004, Kaspi2007, Denney2010, Xiao2011, Grier2012, Bentz2013, Barth2015, Edelson2015, Shen2015,  Fausnaugh2016, Grier2017, Du2018, Lu2022, Pandey2022, Chen2023, Malik2023, Zastrocky2024}. Some AGNs are extremely variable on scales of months or years. These highly variable AGNs may be recognized as changing-look AGNs (CL-AGNs) and challenge our understanding of the SMBH accretion \citep[e.g.,][]{LaMassa2015, Yang2018, MacLeod2019, Guo2020, Wang2024a, Zhu2024}.

Although many previous works studied AGN UV/optical variability \citep[e.g.,][]{MacLeod2010,Gardner2017,Lawrence2018,Hagen2024,Kammoun2024}, the physical mechanisms that cause AGN UV/optical variations are not yet fully clear. A number of theoretical models are proposed. For instance, \cite{Dexter2011} proposed the strongly inhomogeneous quasar accretion disk model and assumed that the temperatures vary independently around the corresponding effective temperature of the standard thin accretion disk \citep[hereafter SSD;][]{Novikov1973, Shakura1973} in each independent disk zone. This model can explain the observed half-light radii from quasar microlensing observations and the observed variability amplitude. \cite{Cai2018} proposed that the local independent temperature fluctuations are affected by the large-scale temperature fluctuation. This model can reproduce the inter-band correlation and time lag of the UV/optical bands and the timescale-dependent bluer-when-brighter color variation. \cite{Krolik1991} proposed the X-ray reprocessing model (i.e., the lamppost model) to explain the time lags of multi-band light curves of AGNs. The X-ray reprocessing model has been adopted to explain continuum RM \citep[e.g.,][]{Cameron2012, Shappee2014, Noda2016, Pahari2020, Fian2024}. \cite{Sun2020} proposed the Corona Heated Accretion disk Reprocessing (CHAR) model to describe the AGN UV/optical variability. The CHAR model can yield the dependence of UV/optical variability on AGN luminosity, black hole mass, and wavelength. 

The light curves of quasars can be studied by spectral techniques that include power spectra \citep[e.g.,][]{Kelly2009}. The stochastic variability (i.e., the light curves) of quasars can be well fitted by the damped random walk (DRW) model \citep[e.g.,][]{Kelly2009, Kozlowski2010, MacLeod2010, Dexter2011, Zu2013, Suberlak2021}, albeit some optical light curves of AGNs diverge from this model \citep[e.g.,][]{Mushotzky2011,Graham2014,Kelly2014}. According to the DRW model, the power spectral density (PSD) of an optical light curve is well described as a power law function $P(f) = f^{-2}$ at the high-frequency end, and white noise at the low-frequency end. The transition frequency between the high-frequency end and the low-frequency end is $f_{0} = 1/(2\pi\tau)$, where $\tau$ represents the theoretical/intrinsic damping timescale that characterizes the light curve. Alternative models \citep[e.g., the damped harmonic oscillator, a.k.a., DHO;][]{Moreno2019} are proposed to fit quasar UV/optical light curves. Given the data quality, it is often true that the model parameters cannot be simultaneously well constrained in models that are more complicated than the DRW model. Furthermore, the damping timescale of the DRW model is related to the timescales of the DHO model \citep[e.g.,][]{Yu2022}. Hence, there are still observational efforts desired to measure the damping timescale. 

Several previous works obtained the relationships between the theoretical/intrinsic damping timescale ($\tau$) in the rest frame and several physical parameters of AGNs, including the wavelength ($\lambda$), monochromatic luminosity ($L_{\mathrm{\lambda}}$), and SMBH mass ($M_{\mathrm{BH}}$). \cite{Kelly2009} analyzed 100 quasars' optical light curves, and found that,
\begin{equation}
\begin{aligned}
    \frac{\tau}{\mathrm{days}} = &(80.4_{-35.8}^{+66.9})(\frac{\lambda L_{\lambda }}{10^{45}\, \mathrm{erg\, s^{-1} }})^{-0.42\pm0.28} \\ &\times (\frac{M_{\mathrm{BH}}}{10^{8} M_{\odot} })^{1.03\pm0.38}.
    \label{equ:Kelly2009}
\end{aligned}
\end{equation}
\cite{MacLeod2010} obtained $\tau$ for $\sim$9000 spectroscopically confirmed quasars' 10-year light curves in Sloan Digital Sky Survey (SDSS) Stripe 82 and reported the relationship between $\tau$ and $\lambda$, and $M_{\mathrm{BH}}$, i.e., $\tau \propto \lambda^{0.17}M_{\mathrm{BH}}^{0.21}$. \cite{Suberlak2021} analyzed 9248 quasars' 15-year light curves by combining the Panoramic Survey Telescope and Rapid Response System 1 Survey data and SDSS Stripe 82 light curves, and reported the correlations between $\tau$ and several AGN physical parameters, including $M_{\mathrm{BH}}$, $\lambda$, and the absolute $i$-band magnitude ($M_{i}$), i.e., $\log_{10}(\tau/\mathrm{days} )= 2.59_{-0.02}^{+0.02} + 0.17_{-0.02}^{+0.02} \mathrm{log_{10}}(\lambda_{\mathrm{rest}}/4000\mathrm{\AA}) + 0.035_{-0.007}^{+0.007} (M_{i} + 23) + 0.14_{-0.02}^{+0.02} \mathrm{log_{10}}(M_{\mathrm{BH}}/10^{9}M_{\odot}) $. They stressed that the DRW parameters could be constrained better by a longer baseline. \cite{Burke2021} measured the damping timescale of 67 AGNs, and obtained the relationship between $\tau$ and $M_{\mathrm{BH}}$, i.e., $\tau = 107_{-12}^{+11}\, \mathrm{days}\, (\frac{M_{\mathrm{BH}}}{10^{8}M_{\odot}})^{0.38_{-0.04}^{+0.05}} $. They proposed that the AGN SMBH mass can be estimated using this relation. \cite{Stone2022} studied the optical variability of 190 quasars in SSDS Stripe 82 with 20-year photometric light curves, and obtained a weak wavelength dependence of $\tau \propto \lambda^{0.2}$. \cite{Zhang2024} obtained $\tau$ for 34 blazars and 7 microquasars from the Fermi-Large Area Telescope and the XMM-Newton X-ray telescope, respectively. They obtained the relation between $\tau$ and $M_{\mathrm{BH}}$, i.e., $\tau = 120_{-18}^{+15}\, \mathrm{days}\, (\frac{M_{\mathrm{BH}}}{10^{8}M_{\odot}})^{0.57_{-0.02}^{+0.02}} $. 

The CHAR model \citep{Sun2020} can be used to predict the relationship between $\tau$ and AGN properties. Indeed, \cite{Zhou2024} investigated the damping timescale in the CHAR model and predicted a new relation between the theoretical/intrinsic damping timescale and the AGN properties, i.e., 
\begin{equation}
\begin{aligned}
    \log_{10}(\tau / \mathrm{days}) =& 0.65\log_{10}(M_{\mathrm{BH}}/M_{\odot})+0.65\log_{10}\dot{m}\\
    &+1.19\log_{10}(\lambda_{\mathrm{rest}}/\mathrm{\AA})-6.04,
\label{equ:Zhou2024-1}
\end{aligned}
\end{equation}
where $\dot{m}$ represents the Eddington ratio (i.e., the ratio of the bolometric luminosity to the Eddington luminosity). Why are the relations obtained from theory and observation inconsistent? Is the discrepancy real or caused by some systematic bias? 

The length of a light curve (i.e., the baseline) must be much longer than the theoretical/intrinsic damping timescale when fitting the light curve with the DRW model. Otherwise, the damping timescale obtained from an insufficiently long light curve will be strongly underestimated \citep[e.g.,][]{Kozlowski2017,Suberlak2021,Hu2024}. \cite{Zhou2024} emphasized that the measured damping timescale can still be significantly biased even if it is less than $10\%$ of the baseline; note that this requirement is often adopted in observational studies; hence, these observational studies still obtained biased damping timescales. Instead, the correct criterion to ensure that the measured timescale is unbiased is that the intrinsic/expected (rather than measured) damping timescale is less than $10\%$ of the baseline. 

In this paper, we constructed unbiased samples based on the relationships between the theoretical/intrinsic damping timescale and the AGN properties obtained from \cite{Kelly2009} (i.e., Equation (\ref{equ:Kelly2009})) and \cite{Zhou2024} (i.e., Equation (\ref{equ:Zhou2024-1})). We stress that all sources in our samples satisfy the criterion that the theoretical/intrinsic damping timescale (rather than the measured timescale) is less than $10\%$ of the baseline. Hence, the measured damping timescales should be unbiased if the relationship of \cite{Kelly2009} or \cite{Zhou2024} is correct. Then, we can use the measured damping timescale to test the two works critically. This manuscript is formatted as follows. In Section \ref{sec:sample}, we describe the sample construction procedures. In Section \ref{sec:data analysis}, we introduce the data analysis. In Section \ref{sec:result}, we present the results. In Section \ref{sec:discussion}, we discuss the possible implications of our results. Conclusions are made in Section \ref{sec:conclusion}.

\section{Observations and sample selection  } \label{sec:sample}

In this work, the quasar optical light curves are obtained from the Zwicky Transient Facility\footnote{\url{https://www.ztf.caltech.edu/}} \citep[ZTF;][]{Bellm2019, Graham2019, Masci2019} and several references. Quasar physical parameters (e.g., black hole mass, luminosity, and redshift) are obtained from the second data release of the $Swift$ BAT AGN Spectroscopic Survey\footnote{\url{ https://www.bass-survey.com/}} \citep[BASS DR2;][]{Koss2022a,Koss2022b}. 

\subsection{Physical parameters} 
\label{sec:properties}

The initial sample is obtained from the BASS DR2 \citep[][]{Koss2022b} which provide the 858 hard-X-ray-selected AGNs in the $Swift$ BAT 70-month sample. This sample provides the key properties for each AGN, e.g., R.A. ($\alpha_{\mathrm{J2000}}$) and Declination. ($\delta_{\mathrm{J2000}}$); AGN type defined by optical spectroscopy, including Sy1 (broad $\mathrm{H\beta}$ emission line), Sy1.9 (narrow $\mathrm{H\beta}$ and broad $\mathrm{H\alpha}$ emission lines), and Sy2 (narrow $\mathrm{H\beta}$ and $\mathrm{H\alpha}$ emission lines). For most of Sy1 AGNs, the redshifts ($z$) are measured by [O{\sc iii}] $\lambda5007$ or broad emission lines \citep[e.g., broad Mg\,{\sc ii} and C\,{\sc iv}; see][]{Koss2022b}. The redshifts of narrow-line sources are obtained by the emission-line [O{\sc iii}] $\lambda5007$ \citep[e.g.,][]{Koss2022b}; the majority of black hole masses are calculated by either broad $\mathrm{H\beta}$ or stellar velocity dispersion measurements. The spectra used for the above measurements were observed by the Palomar Hale 5 m telescope Double Beam Spectrograph (DBSP) or the Very Large Telescope (VLT) X-shooter spectrograph \citep[e.g.,][]{Koss2022a}. The bolometric luminosity ($L_{\mathrm{bol}}$) is calculated by the 14–150 keV intrinsic luminosity with the bolometric correction of 8, and assuming the photon index $\Gamma$ = 1.8. In this work, we focus on Seyfert 1 AGNs (``Type'' = Sy1). There are 359 Seyfert 1 AGNs out of the 858 hard-X-ray-selected AGNs.

\subsection{Light curves}
The light curves of the 359 Seyfert 1 AGNs are obtained from ZTF. ZTF is a time-domain survey using the 48-inch Samuel Oschin Telescope of Palomar Observatory. The entire northern visible sky has been scanned at the optical band (including three custom filters, $zg$, $zr$, and $zi$) by the ZTF camera every 2 days since 2018, and the camera has a 47 deg$^{2}$ field with 600 megapixels. The effective wavelengths of the $zg$, $zr$, and $zi$ bands are 4753, 6369, and 7915 $\mathrm{\AA}$, respectively \citep[][]{Rodrigo2012, Rodrigo2020}. The ZTF data are released every 2 months, and the baselines of current $g$ and $r$ band light curves are $\sim$ 1800 days. Such high-quality ZTF light curves, including the long baseline and the high cadence, are very suitable for time-domain science, such as studying quasar optical variability. The $zi$ band light curves are not utilized in this work because their baselines are not long enough for our research. In this work, we retrieved the $zg$ and $zr$ light curves for each source in our sample from ZTF DR21 via the IRSA service\footnote{\url{https://irsa.ipac.caltech.edu/frontpage/}}, and the cross-matching radius is $1''$. The observations are obtained from all CCDs. For each light curve, we removed observations with poor quality (i.e., ``catflags''$>$ 0). 

\subsection{Sample selection}

As emphasized by several previous works \citep[e.g.,][]{Kozlowski2017,Suberlak2021,Kozlowski2021,Hu2024,Zhou2024}, the damping timescale is significantly biased if the light-curve baseline is less than ten times of the theoretical damping timescale. As we have stressed in Section \ref{sec:intro}, the commonly adopted criterion, which is that the baseline is at least ten times longer than the best-fitting damping timescale, cannot eliminate this bias \citep[also see][]{Zhou2024}. Hence, in the following two sections, we aim to construct samples according to our new criterion, i.e., the theoretical damping timescale should be less than $10\%$ of the baseline. As a result, the sample selection depends critically upon the damping timescale relation we aim to test in this work.

\subsubsection{Sample selection based on \cite{Kelly2009}} \label{sec:sample1}
In this section, we aim to construct a sample to test the damping timescale relation of \cite{Kelly2009}, i.e., Equation \ref{equ:Kelly2009}. Given that the rest-frame baseline for ZTF light curves is about $1800/(1+z)$ days ( where $z$ represents the redshift), we expect that the measured damping timescale can be unbiased if the \textit{theoretical damping timescale} is less than $180/(1+z)$ days (i.e., less than $10\%$ of the baseline). This requirement puts a strong constraint on $M_{\mathrm{BH}}$, luminosity, and redshift. The black hole mass of each source can be obtained from the BASS DR2 (see the description in Section \ref{sec:properties}). Sources without mass estimates (``logMBH'' = 0) are removed. The monochromatic luminosity at 5100 $\mathrm{\AA}$ ($L_{\mathrm{5100}}$) of each source can be calculated as follows, i.e., $L_{5100}=L_{\mathrm{bol}}/9.26$ \citep[e.g.,][]{Richards2006,Shen2011}. For each source, we use $M_{\mathrm{BH}}$,  $L_{\mathrm{bol}}$, and Equation \ref{equ:Kelly2009} to calculate the theoretical damping timescale ($\tau_{\mathrm{th,Kelly+2009}}$). Only sources with $\tau_{\mathrm{th,Kelly+2009}}<180/(1+z)$ days are selected. There are 142 sources in the $zg$ band and 142 sources in the $zr$ band of the 359 Seyfert 1 that merit this requirement and are selected for subsequent analysis, respectively. 

\subsubsection{Sample selection based on \cite{Zhou2024}} 

\cite{Zhou2024} reported the relationship between $\tau$ and AGN properties (i.e., $M_{\mathrm{BH}}$, $\dot m$, and $\lambda_{\mathrm{rest}}$), i.e., Equation (\ref{equ:Zhou2024-1}). The dimensionless accretion ratio $\dot{m} = L_{\mathrm{bol}}/((1+k)\times L_{\mathrm{edd}})$, $k = \frac{1}{3}$ represents the ratio between the magnetic fluctuations' power of the corona and the dissipation rate of the disk turbulent magnetic power \citep[for details, see][]{Sun2020}, where $L_{\mathrm{edd}} = 1.26 \times 10^{38} \times (M_{\mathrm{BH}}/M_{\odot})\,  \mathrm{erg \,s^{-1}}$ is the Eddington luminosity. Hence, Equation (\ref{equ:Zhou2024-1}) can be expressed in the form of
\begin{equation} \label{equ:Zhou2024-2}
\begin{aligned}
    \log_{10}(\tau / \mathrm{days}) =\, &0.65\log_{10}(L_{\mathrm{bol}}/\mathrm{erg\, s^{-1} }) \\ &+ 1.19\log_{10}(\lambda_{\mathrm{rest}}/\mathrm{\AA})-30.9.
\end{aligned}
\end{equation}

We calculate the theoretical damping timescale ($\tau_{\mathrm{th,Zhou+2024}}$) of \cite{Zhou2024} for each of the $359$ Seyfert 1 AGNs in BASS DR2. Then, we again only select sources with $\tau_{\mathrm{th,Zhou+2024}}<180/(1+z)$ days. There are 70 and 38 sources of the 359 Seyfert 1 that meet our requirement for the $zg$ band and $zr$ band, respectively.

\subsection{Other sources obtained from the literature}
Several narrow-line Seyfert 1 galaxies' light curves have been analyzed in the literature. The $Kepler$ light curve ($\lambda = 6600 \mathrm{\AA}$) of Zw 229-15 has been analyzed by \cite{Chen2015}. The black hole mass of Zw 229-15 is $\mathrm{log_{10}}(M_{\mathrm{BH}} / M_{\odot}) = 6.91$ and the redshift is $z = 0.0275$ \citep[][]{Koss2022b}. The $Swift$ UVOT $UVM2$ band $(\lambda = 2600\mathrm{\AA})$ light curve of WPVS 007 has been analyzed by \cite{Li2019}, whose $M_{\mathrm{BH}} = 4.1\times10^{6}\, M_{\odot}$ \citep[][]{Leighly2015}, and $z = 0.02882$ \citep[][]{Grupe1995}. The $Swift$ UVOT $UVW2$ band $(\lambda = 2120\mathrm{\AA})$ light curve of Mrk 335 has been analyzed by \cite{Griffiths2021}; the black hole mass and redshift of Mrk 335 are $\mathrm{log_{10}}(M_{\mathrm{BH}} / M_{\odot}) = 7.32$ and $z = 0.0259$, respectively \citep[][]{Koss2022b}. \cite{Burke2021} reported the characteristic optical variability timescale of 67 AGNs, and 6 sources (including NGC 4395, NGC 5548, DES J021822.51-043036.0, SDSS J025007.03+002525.3, SDSS J153425.58+040806.7, and SDSS J160531.85+174826.3) of 67 AGNs satisfy the condition $\tau_{\mathrm{th,Kelly+2009}}/\mathrm{baseline}<10\%$, and $\tau_{\mathrm{th,Zhou+2024}}/\mathrm{baseline}<10\%$. Five sources were added to our sample; NGC 4395 has been reduced since this target is included in our Swift BAT sample. \cite{Zhou2024} measured the damping timescale for 3 local AGNs \citep[see Table 5 in][]{Zhou2024}, including NGC 4151, NGC 7469, and NGC 3516. These sources were added to our sample. We confirm that our main conclusions remain unchanged if these sources were excluded in our subsequent analysis. 

\begin{figure*}[ht!]
\plotone{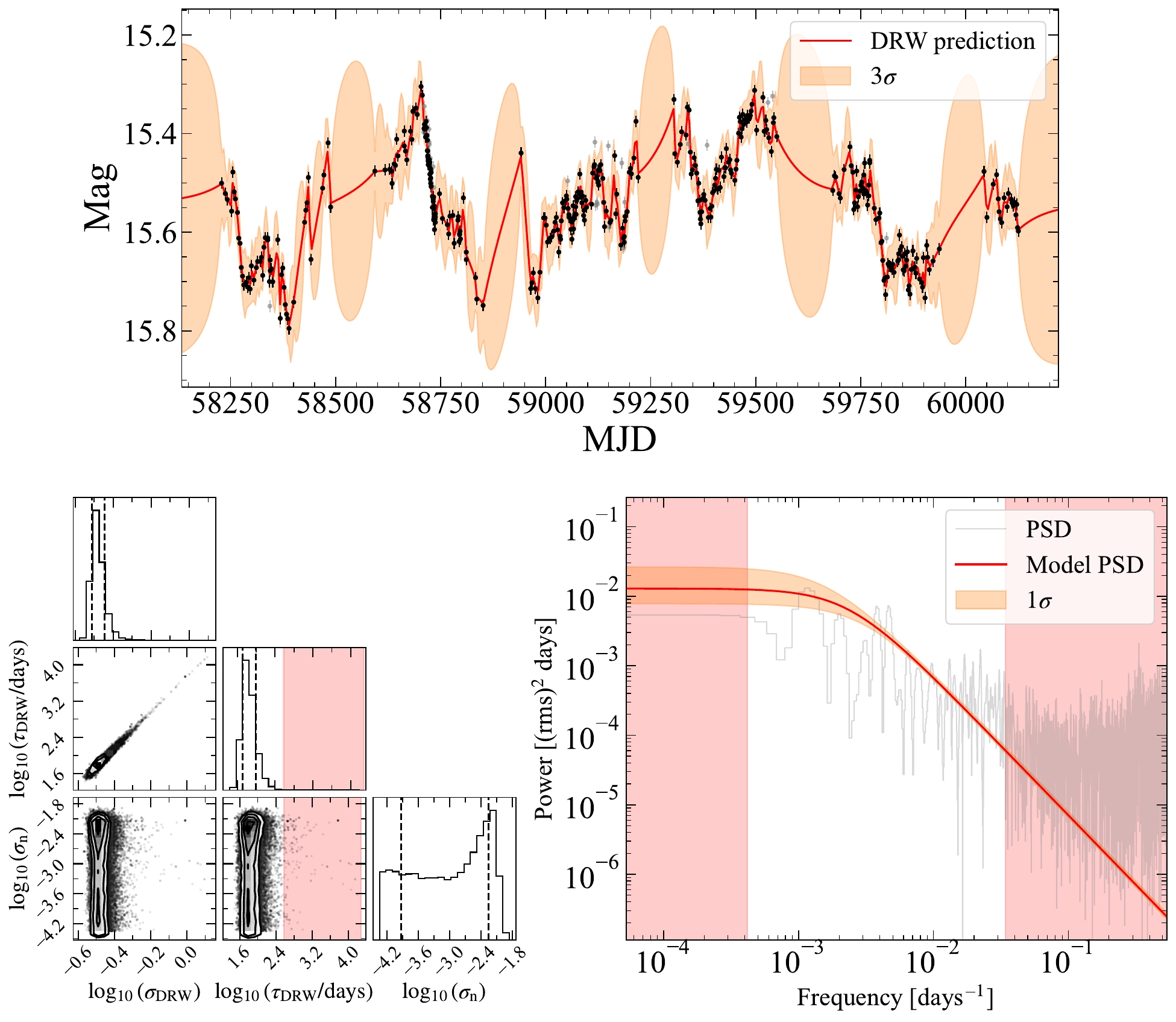}
\caption{Fitting the $zg$ band light curve (in the observed-frame) of Z 493-2 with the DRW model via the \textbf{\texttt{taufit}} code. Top panel: the solid red curve and shaded regions are the best-fitting DRW model and the $3\sigma$ confidence intervals. The gray and black data points correspond to the observations that lie outside or inside the shaded regions, respectively. Bottom-left panel: the posterior distributions of the DRW parameters obtained in our new DRW fit (i.e., the DRW fit with the black data points in the top panel). The parameter $\sigma_{\mathrm{DRW}}$ is the long-term variability amplitude, $\tau_{\mathrm{DRW}}$ is the damping timescale, and $\sigma_{\mathrm{n}}$ is the excess white noise amplitude. The medians of the posterior distributions are adopted as the best-fitting parameters, and the 1$\sigma$ uncertainties are obtained by the 16-th and 84-th percentiles of the posterior distributions. The red shaded regions in the bottom-left panel correspond to where $\tau_{\mathrm{DRW}}$ is greater than 10$\%$ of the baseline. Bottom-right panel: the gray curve is the PSD of the data (the black data points in the top panel), and the DRW PSD (the red solid curve) and its 1$\sigma$ uncertainties (the orange shaded regions). The red-shaded regions represent the regions of frequency space not sampled by the light curve, including the regions with timescales shorter than the minimum cadence and the timescales longer than 20$\%$ of the baseline.
\label{Fig1}}
\end{figure*}

\begin{figure*}[ht!]
\plotone{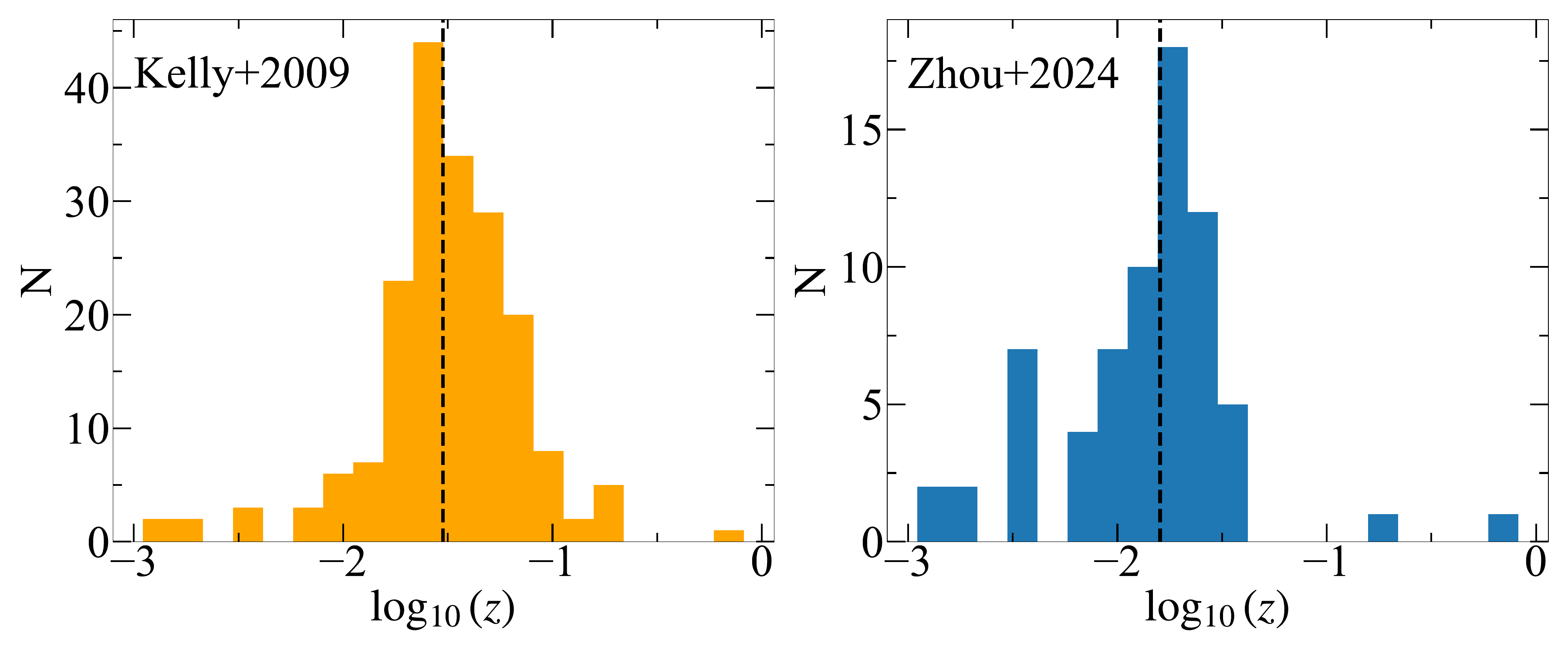}
\caption{The redshift distributions of the samples in this work. Left panel: the selected sources with theoretical damping timescales, which are calculated from Equation \ref{equ:Kelly2009} \citep{Kelly2009}, smaller than $10\%$ of the light curve durations. The redshift ranges from 0.001 to 0.823, and the median redshift is 0.030 (the black dashed line). Right panel: the selected sources with theoretical damping timescales, which are calculated from Equation \ref{equ:Zhou2024-1} \citep{Zhou2024}, smaller than $10\%$ of the light curve durations. The redshift ranges from 0.001 to 0.823, and the median redshift is 0.016 (the black dashed line). 
\label{Fig2}}
\end{figure*}

\begin{figure}[ht!]
\epsscale{1.3}
\plotone{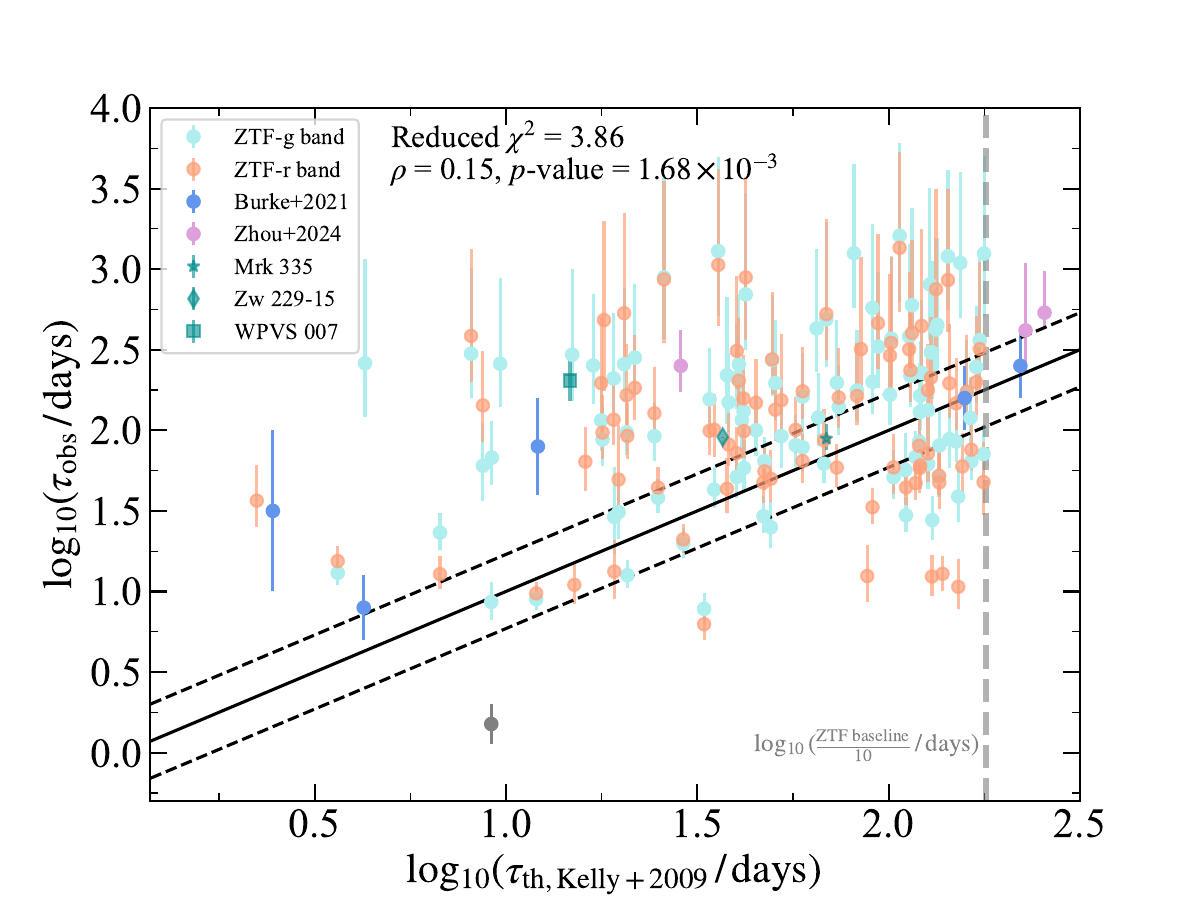}
\caption{The relationship between the damping timescales ($\log_{10}(\tau_{\mathrm{obs}})$; rest-frame values) obtained from the DRW fittings and the theoretical predictions ($\log_{10}(\tau_{\mathrm{th, Kelly+2009}})$) of \cite{Kelly2009} (i.e., Equation \ref{equ:Kelly2009}). The cyan and orange solid dots indicate the damping timescales obtained by the $zg$ and $zr$ band light curves, respectively. The blue and purple dots represent the damping timescales obtained from \cite{Burke2021} and \cite{Zhou2024}, respectively. The dark green star, diamond, and square indicate the damping timescales of Mrk 335, Zw 229-15, and WPVS 007, respectively. For more information on these sources, see Table \ref{Tab1}. The gray point (ESO 424-12 $zr$ band) was not considered in our analysis because of its unusually small measured damping timescale (the DRW fitting result of ESO 424-12 $zr$ band light curve is shown in Figu.~\ref{FigB1}). The vertical solid gray line represents 10$\%$ of the baseline of the ZTF light curve. The solid black line indicates a one-to-one line, and the black dashed line indicates the uncertainties of 0.2 dex. The reduced $\chi^{2}$ between $\log_{10}(\tau_{\mathrm{th, Kelly+2009}})$ and $\log_{10}(\tau_{\mathrm{obs}})$ is 3.86, and the Kendall's correlation coefficient is $\rho$ = 0.15 ($p$-value = $1.68\times10^{-3}$). Most of the data points are located above the one-to-one line. That is, $\log_{10}(\tau_{\mathrm{th, Kelly+2009}})$ significantly underestimates the observed damping timescales. 
\label{Fig3}}
\end{figure}

\begin{figure}[ht!]
\epsscale{1.3}
\plotone{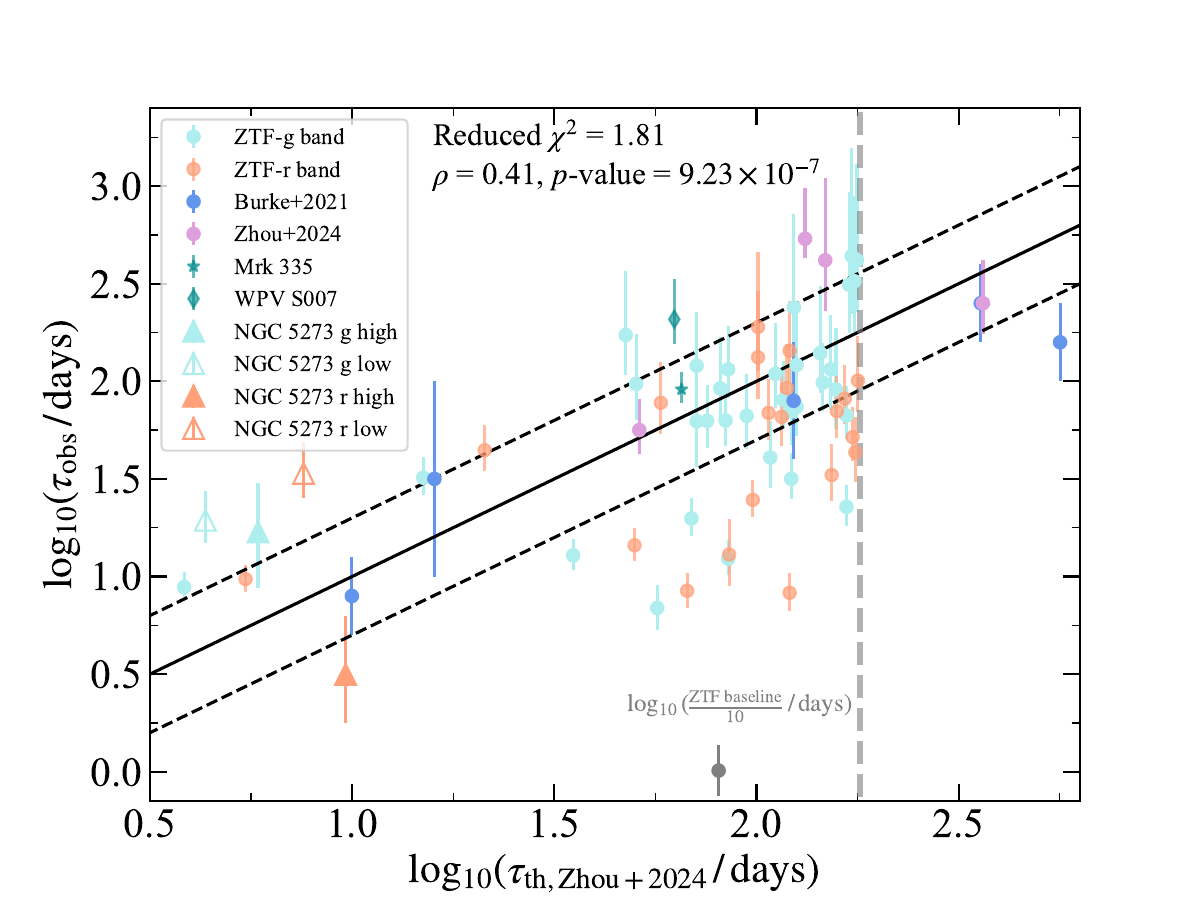}
\caption{The relationship between the damping timescales ($\log_{10}(\tau_{\mathrm{obs}})$; rest-frame values) obtained from DRW fitting and the theoretical predictions ($\log_{10}(\tau_{\mathrm{th, Zhou+2024}})$) of \cite{Zhou2024} (i.e., Equation \ref{equ:Zhou2024-2}). The meaning of the data points and lines in this figure is the same as in Fig.~\ref{Fig3}. The triangles represent the damping timescales of the high and low states of $zg$ and $zr$ of the CL AGN NGC 5273, respectively. For more information on these sources, see Table \ref{Tab2}. The reduced $\chi^{2}$ between $\log_{10}(\tau_{\mathrm{th, Zhou+2024}})$ and $\log_{10}(\tau_{\mathrm{obs}})$ is 1.81, and the Kendall's correlation coefficient is $\rho$ = 0.41 ($p$-value = $9.23\times10^{-7}$). The observations and the theoretical predictions of \cite{Zhou2024} are roughly consistent.
\label{Fig4}}
\end{figure}

\begin{figure*}[ht!]
\epsscale{1.2}
\plotone{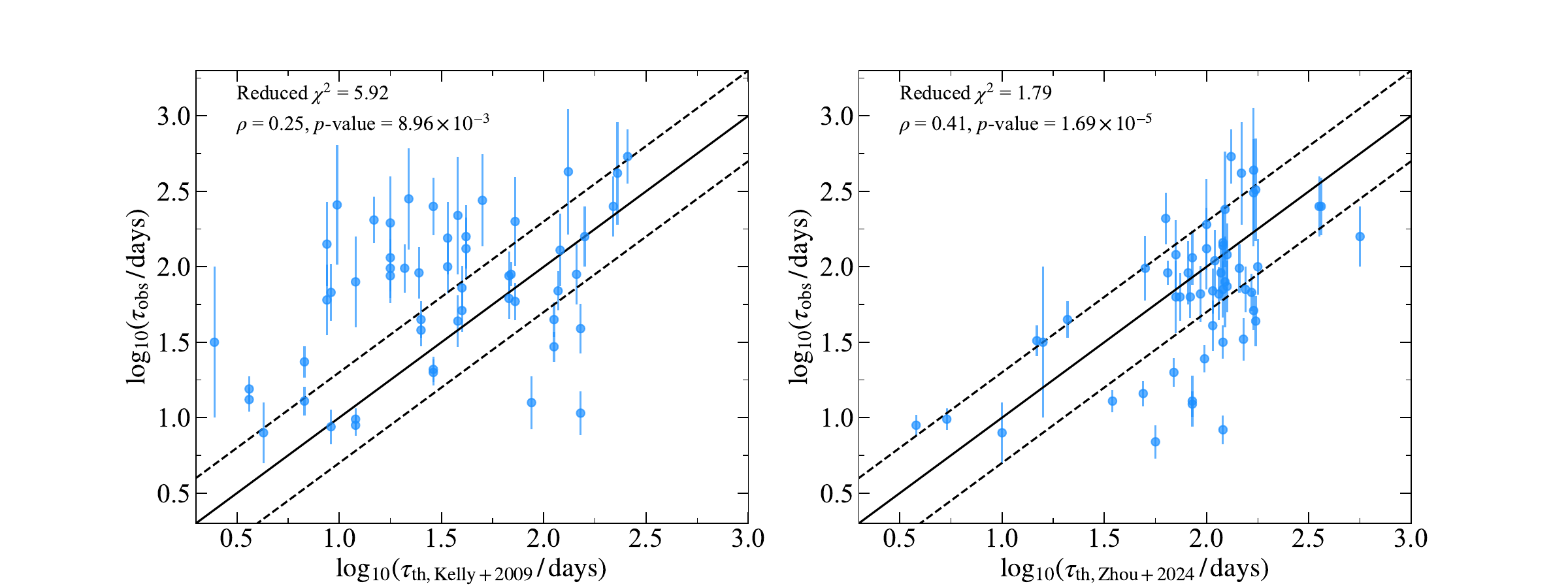}
\caption{Left panel: the same as Fig.~\ref{Fig3} but for sample C, which consists of sources that satisfy the criteria according to \cite{Kelly2009} (i.e., Equation \ref{equ:Kelly2009}) and \cite{Zhou2024} (i.e., Equation \ref{equ:Zhou2024-2}). Right panel: the same as Fig.~\ref{Fig4} but for sample C. Again, observations are more consistent with the theoretical predictions of \cite{Zhou2024} than those of \cite{Kelly2009}. 
\label{Fig5}}
\end{figure*}

\begin{figure}[ht!]
\epsscale{1.3}
\plotone{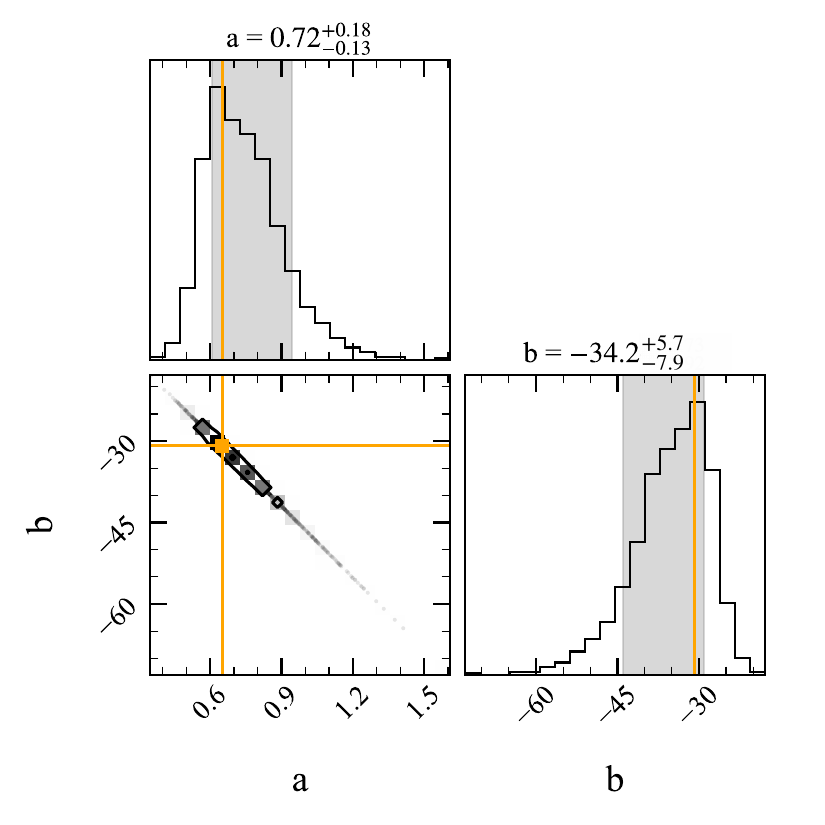}
\caption{The probability distributions of the parameters $a$ and $b$ in the fitting relation $\log_{10}(\tau_{\mathrm{obs}}) = a \log_{10}(L_{\mathrm{bol}}) + 1.19 \log_{10}(\lambda_{\mathrm{rest}}) + b$. Bottom-left panel: the joint probability distributions of $a$ and $b$. Top-left panel: the distribution of $a$. The median value of the distribution is adopted as the best-fitting $a$, and the gray region is the 1$\sigma$ uncertainties of $a$. The vertical solid orange line indicates the same parameter predicted by \cite{Zhou2024}, which is $0.65$. Bottom-right panel: the distribution of $b$. The median value of the distribution is adopted as the best-fitting $b$, and the gray region is the 1$\sigma$ uncertainties of $b$. The vertical dashed orange line represents the same parameter predicted by \cite{Zhou2024}, which is $-30.9$. Our best-fitting parameters and the theoretical predictions of \cite{Zhou2024} are statistically consistent. 
\label{Fig6}}
\end{figure}

\section{Data analysis: light curve fitting \label{sec:data analysis}}

AGN light curves can be well fitted by the DRW model based on the \textbf{\texttt{taufit}}\footnote{\url{https://github.com/burke86/taufit}} code \citep[][]{Burke2021}. This code is built upon the fast Gaussian process solver \textbf{\texttt{celerite}} \citep{Foreman2017}. \textbf{\texttt{celerite}} fits the time series to the following specified kernel function using the Gaussian process regression

\begin{equation}
    k(t_{nm}) = 2\sigma^{2}_{\mathrm{DRW}}\mathrm{exp}(-t_{nm}/\tau_{\mathrm{DRW}})+\sigma^{2}_{n}\delta_{nm},
\end{equation}
where $t_{nm} = \left | t_{n} - t_{m} \right | $ represents the time interval between two measurements ($t_{n}, t_{m}$) of the light curve. $\sigma_{\mathrm{DRW}}$ represents the long-term standard deviation of variability. $\tau_{\mathrm{DRW}}$ defines the damping timescale. $\sigma_{\mathrm{n}}$ is the excess white noise amplitude, which is added to the DRW kernel to account for the possible underestimation of the photometric uncertainties. $\sigma^{2}_{n}$ represents the variance of the noise component. $\delta_{nm}$ is the Kronecker delta function. The prior distribution of $\tau_{\mathrm{DRW}}$ was set to be a uniform distribution whose lower and upper bounds are the minimum cadence to the 10 times the baseline of each light curve, respectively. The best-fitting $\tau_{\mathrm{DRW}}$ is obtained by maximizing the posterior probability. The joint posterior probability distributions of the DRW parameters from \textbf{\texttt{celerite}} are obtained by the Markov Chain Monte Carlo (MCMC) code sampler \textbf{\texttt{emcee}} \citep[][]{Foreman-Mackey2013} implemented in Python. The median values of the posterior distributions are adopted as the best-fitting parameters, and 16-th and 84-th percentiles of the posterior distributions are adopted as the 1$\sigma$ uncertainties for the best-fitting parameters. We compare the Akaike Information Criterion \citep[AIC;][]{Akaike1998} of this fit ($\mathrm{AIC}_{\mathrm{best}}$) with the AICs of two alternative fits. The first alternative fit ($\mathrm{AIC}_{\mathrm{low}}$) is that we set $\tau_{\mathrm{DRW}}$ to be an extremely small value ($0.02$ days, i.e., about one-hundredth of the typical ZTF cadence); the second ($\mathrm{AIC}_{\mathrm{up}}$) is that we set $\tau_{\mathrm{DRW}}$ to be an extremely large value ($18000$ days, i.e., about 100 times of the typical ZTF baseline). The AIC of each fit is calculated as $-\ln(\mathrm{Likelihood})+2N$, where $N=3$ is the number of parameters. Statistically speaking, if the AIC difference between two models is larger than $10$, the model with a larger AIC value has relatively little support \citep[e.g.,][]{Burnham2011}. Hence, we propose that the damping timescale is well constrained only if $\Delta \mathrm{AIC_{up}} = \mathrm{AIC_{up}} - \mathrm{AIC_{best}} > 10$ and $\Delta \mathrm{AIC_{low}} = \mathrm{AIC_{low}} - \mathrm{AIC_{best}} > 100$ (the larger $\Delta \mathrm{AIC}$ corresponds to the better constraint of the damping timescale).

In this section, the light curve(in the observed-frame) of each source mentioned in Section \ref{sec:sample} is fitted by the above method. A DRW fit example of Z 493-2 is shown in Fig. \ref{Fig1}. The top panel shows the original light curve of Z 493-2, and the predicted light curve from the DRW model obtained by the best-fitting parameters. Note that, we removed the observations outside the 3$\sigma$ of the predicted light curve (the gray points in the top panel), and then repeated the fitting process until all observations were included within 3$\sigma$ of the predicted light curve obtained by the DRW model. The bottom left panel shows the best-fitting DRW parameters (i.e., $\sigma_{\mathrm{DRW}}$, $\tau_{\mathrm{DRW}}$, and $\sigma_{\mathrm{n}}$) obtained by the \textbf{\texttt{taufit}}. The bottom right panel shows the PSD of the original light curve (the gray curve), and the PSD of the light curve obtained from the posterior distribution of the \textbf{\texttt{celerite}} fit (the red curve). The model PSD is shown in the orange region (with 1$\sigma$ uncertainty).

In summary, our sample selection criteria are summarized as follows.

\begin{itemize}
\item The theoretical damping timescale obtained from \cite{Kelly2009} (i.e., $\tau_{\mathrm{th, Kelly+2009}}$ calculated by Equation \ref{equ:Kelly2009}) or \cite{Zhou2024} (i.e., $\tau_{\mathrm{th,Zhou+2024}}$ calculated by Equation \ref{equ:Zhou2024-2}) should be smaller than the 10$\%$ baseline in the rest frame, i.e., $\tau_{\mathrm{th, Kelly+2009}} < 10\% \times \mathrm{baseline}/(1+z)$, or $\tau_{\mathrm{th,Zhou+2024}} < 10\% \times \mathrm{baseline}/(1+z)$;
\item Following \cite{Burke2021}, the best-fitting damping timescale obtained by the DRW model fitting ($\tau_{\mathrm{obs}}$) is larger than 1.5 times cadence, i.e., $\tau_{\mathrm{obs}} > 1.5 \times \mathrm{cadence}$;
\item Following \cite{Burke2021}, the source should have the high signal-to-noise ratio, i.e., $\sigma^{2}_{\mathrm{DRW}} > \sigma^{2}_{\mathrm{n}} + \mathrm{d}y^{2}$, where $\mathrm{d}y$ is the median magnitude err of each light curve;
\item As described in Section \ref{sec:data analysis}, in order to ensure that the damping timescales of the sources in our sample are indeed constrained by the data, we require that $\Delta \mathrm{AIC_{low}} > 100$, and $\Delta \mathrm{AIC_{up}} > 10$.
\end{itemize}

The key parameters of each source in our sample include baseline, cadence (the median cadence of each light curve), $\tau_{\mathrm{obs}}$ (i.e., obtained by the DRW fit), $\sigma_{\mathrm{DRW}}$, $\sigma_{\mathrm{n}}$, $L_{\mathrm{bol}}$, $M_{\mathrm{BH}}$, $\lambda$, and $z$ (see Tables \ref{Tab1} and \ref{Tab2}). The theoretical damping timescale ($\tau_{\mathrm{th, Kelly+2009}}$ and $\tau_{\mathrm{th, Zhou+2024}}$) of our sample can be calculated by Equation \ref{equ:Kelly2009} and Equation \ref{equ:Zhou2024-2}. Based on the damping timescale predicted by \cite{Kelly2009} (i.e., Equation \ref{equ:Kelly2009}) and the above criteria, 112 sources (including 187 light curves) were included in our final sample (hereafter sample K) to test the \cite{Kelly2009} relation. More details of these sources are listed in Table \ref{Tab1}. Meanwhile, 49 sources (including 68 light curves) were included in our final sample (hereafter sample Z) based on the damping timescale predicted by \cite{Zhou2024}; i.e., Equation \ref{equ:Zhou2024-2} and the above conditions. More details of these sources are listed in Table \ref{Tab2}. The redshift distribution of these sources ranges from 0.001 to 0.823 (see Fig. \ref{Fig2}), with medians of 0.030 and 0.016, respectively. All sources in the sample are low redshift AGNs.

\section{Results \label{sec:result}}
\subsection{Observed damping timescales vs. theoretical predictions \label{subsec:4.1}}
The relationship between the damping timescales ($\log_{10}(\tau_{\mathrm{obs}})$) obtained from the DRW fitting and the theoretical predictions ($\log_{10}(\tau_{\mathrm{th,Kelly+2009}})$) of \cite{Kelly2009} (i.e., Equation \ref{equ:Kelly2009}) is shown in Fig. \ref{Fig3}. There are 187 data points: 93 light curves obtained from the $zg$ band (i.e., the cyan solid dots), 83 light curves obtained from the $zr$ band (i.e., the orange solid dots), 5 data points obtained \cite{Burke2021} (i.e., the blue solid dots), 3 data points obtained from \cite{Zhou2024} (i.e., the purple solid dots), and 3 green dots (Mrk 335, Zw 229-15, and WPVS 007) are obtained from the literature \citep[i.e.,][]{Griffiths2021, Chen2015, Li2019}. We found that there are unusually dense observations for the $zr$ band light curve of ESO 424-12, which leads to the damping timescale of this light curve being unusually small, i.e., about 1 day (the DRW fitting result of ESO 424-12 $zr$ band light curve is shown in Fig.~\ref{FigB1}). We reject this light curve from our sample. Note that our conclusion remains unchanged if this target is included. The vertical solid gray line represents $10\%$ of the baseline of the ZTF light curve (i.e., $\mathrm{ZTF\,baseline}/10$ days). It is important to emphasize that all sources in our sample satisfy the theoretically predicted damping timescale being less than $10\%$ of its baseline in the rest frame. As shown in Fig. \ref{Fig3}, the vast majority of data points (184 of 187) lie to the left of the vertical dashed gray line. There are three data points located to the right of the vertical line because their light curves are not obtained from the ZTF. Their light curves have baselines much larger than 1800 days. We calculated the reduced $\chi^2$ between $\log_{10}(\tau_{\mathrm{th,Kelly+2009}})$ and $\log_{10}(\tau_{\mathrm{obs}})$ to be 3.86 and found that the Kendall's correlation coefficient to be $\rho = 0.15$ with a $p$-value of $1.68\times10^{-3}$. The reduced $\chi^2$ is $ \chi^{2}_{\mathrm{reduced} }=\frac{1}{K-n}\sum_{i=1}^{K}\left ((\log_{10}(\tau_{\mathrm{obs}, i})-\log_{10}(\tau_{\mathrm{th,Kelly+2009}, i}))^2/\sigma^{2}_{i}  \right ) = 3.86$, where $K$ is the total number of data points, $n = 0$ is the number of free parameters in Equation \ref{equ:Kelly2009}, $\log_{10}(\tau_{\mathrm{obs}, i})$ is the value at the $i$-th observed damping timescale, $\log_{10}(\tau_{\mathrm{th,Kelly+2009}, i})$ is the value at the $i$-th predicted damping timescale, and $\sigma_{i}$ is the error associated with the $i$-th observed damping timescale. The large reduced $\chi^2$ and small correlation coefficient suggest that the measured timescales cannot be well described by Equation \ref{equ:Kelly2009} \citep{Kelly2009}. Indeed, most of the points are located above the one-to-one line. That is, $\log_{10}(\tau_{\mathrm{th,Kelly+2009}})$ significantly underestimates the observed damping timescales.

Similarly, the relationship between the damping timescales ($\log_{10}(\tau_{\mathrm{obs}})$) obtained from the DRW fitting and the theoretical predictions ($\log_{10}(\tau_{\mathrm{th,Zhou+2024}})$) of \cite{Zhou2024} (i.e., Equation \ref{equ:Zhou2024-2}) is shown in Fig. \ref{Fig4}. There are 68 data points: 35 are obtained from the $zg$ band (i.e., the cyan solid dots), 22 are obtained from the $zr$ band (i.e., the orange solid dots), 5 data points are obtained from \cite{Burke2021} (i.e., the blue solid dots), 4 data points are obtained from \cite{Zhou2024} (i.e., the purple solid dots), and 2 green dots (Mrk 335 and WPVS 007) are obtained from the literature \citep[i.e.,][]{Griffiths2021, Li2019}. NGC 5273 is a CL AGN, and four light curves of the high and low states of this source are studied separately. We again removed ESO 424-12 from our sample. As shown in Fig. \ref{Fig4}, the vast majority of data points (65 of 68) lie to the left of the vertical dashed gray line. There are three data points located to the right of the vertical line because their light curves are not obtained from the ZTF. We calculate the reduced $\chi^2$ between $\log_{10}(\tau_{\mathrm{th,Zhou2024}})$ and $\log_{10}(\tau_{\mathrm{obs}})$ to be 1.81 and the Kendall's correlation coefficient to be $\rho = 0.41$ with a $p$-value of $9.23\times10^{-7}$. This indicates that the measured timescales can be well described by the relation of \cite{Zhou2024}. It can be noticed that over half of the data points (38 of 68) lie within 0.2 dex of the uncertainty of the one-to-one line. That is, the observations and the theoretical predictions of \cite{Zhou2024} are roughly consistent.

We also construct a new sample (hereafter sample C) by selecting sources whose theoretical predictions according to \cite{Kelly2009} (i.e., Equation \ref{equ:Kelly2009}) and \cite{Zhou2024} (i.e., Equation \ref{equ:Zhou2024-2}) satisfy our criteria (see Section~\ref{sec:data analysis}); that is, sample C is an intersection of the sample in Fig.~\ref{Fig3} and the sample in Fig.~\ref{Fig4}. There are 38 sources (including 54 light curves) in sample C. The relationship between the theoretical/intrinsic damping timescales ($\log_{10}(\tau_{\mathrm{obs}})$) obtained from the DRW fittings and the theoretical predictions ($\log_{10}(\tau_{\mathrm{th}})$) of sample C are shown in Fig.~\ref{Fig5}. The reduced $\chi^2$ between $\log_{10}(\tau_{\mathrm{th}})$ of Equation \ref{equ:Kelly2009} and $\log_{10}(\tau_{\mathrm{obs}})$ is $5.92$ (the left panel of Fig.~\ref{Fig5}), and the Kendall's correlation coefficient is $\rho = 0.25$ with a $p$-value of $8.96\times10^{-3}$. There are less than half of the data points (22 of 54) lie within 0.2 dex of the uncertainty of the one-to-one line, and most of the points are located above the one-to-one line. Meanwhile, the reduced $\chi^2$ between $\log_{10}(\tau_{\mathrm{th}})$ of Equation \ref{equ:Zhou2024-2} and $\log_{10}(\tau_{\mathrm{obs}})$ is $1.79$ (the right panel of Fig.~\ref{Fig5}), and the Kendall's correlation coefficient is $\rho = 0.41$ with a $p$-value of $1.69\times10^{-5}$. More than half of the data points (32 of 54) lie within 0.2 dex of the uncertainty of the one-to-one line. In summary, the measured damping timescales can be better described by the relation of \cite{Zhou2024} rather than \cite{Kelly2009}.

\subsection{The dependencies of $\tau_{\mathrm{obs}}$ upon $L_{\mathrm{bol}}$ and $\lambda_{\mathrm{rest}}$}\label{sec:parameters fit}
As described in Section \ref{sec:intro}, some previous works reported the established empirical relations between the rest frame damping timescale and several physical parameters. The empirical relations between $\tau_{\mathrm{obs}}$ and $L_{\mathrm{bol}}$, and $\lambda_{\mathrm{rest}}$ are also constructed in this section. The adopted sample is the same as in Fig. \ref{Fig4} because the observations are more consistent with the theoretical predictions of \cite{Zhou2024} than that of \cite{Kelly2009}. The fitted equation is 
\begin{equation}
\begin{aligned}
    \log_{10}(\frac{\tau_{\mathrm{obs}}}{\mathrm{days}}) = & a \log_{10}(\frac{L_{\mathrm{bol}}}{\mathrm{erg\, s^{-1}}})\\ & + 1.19 \log_{10}(\frac{\lambda_{\mathrm{rest}}}{\mathrm{\AA}}) + b.
\end{aligned}
\end{equation}
The best-fitting parameters are obtained by minimizing the $\chi^2$ statistic. The $\chi^2$ statistic is $\chi^2 = \sum ((\tau_{\mathrm{obs,i}}-\tau_{\mathrm{model,i} })^{2}/((\tau_{\mathrm{err,obs,i} })^{2}+(a\times \delta\log _{10}{L_{\mathrm{bol,i} }})^{2}))$, where $\tau_{\mathrm{err,obs,i} }$ is the 1$\sigma$ uncertainty of $\tau_{\mathrm{obs,i}}$, and $\delta\log _{10}{L_{\mathrm{bol,i} }}$ represents the typical 0.2 dex uncertainty of $\log _{10}{L_{\mathrm{bol} }}$ for each source. We use bootstrap with replacements to obtain the statistical distributions of $a$ and $b$. The median values of the bootstrap distributions are adopted as the best-fitting parameters, and the 1$\sigma$ uncertainties are taken as 16-th to 84-th percentiles of the distributions. The results are shown in Fig. \ref{Fig6}, with the best-fitting parameters with 1$\sigma$ uncertainties being $a = 0.72^{+0.18}_{-0.13}$ and $b = -34.2^{+5.7}_{-7.9}$ (i.e., the histograms and gray shaded areas in the upper left and lower right panels). The solid orange lines indicate the parameters (i.e., $a_{\mathrm{Zhou+2024}} = 0.65$, $b_{\mathrm{Zhou+2024}} = -30.9$) reported by \cite{Zhou2024}. The best-fitting parameters obtained in this work and those of \cite{Zhou2024} are statistically consistent. Note that the parameters of the wavelength term are fixed to $1.19$ according to \cite{Zhou2024}. This is because most of the sources in our sample are from the ZTF and the wavelengths in the $zg$ and the $zr$ bands are close, i.e., $\left (\lambda_{\mathrm{r}}-\lambda_{\mathrm{g}}\right) / \left (\overline{\lambda_{\mathrm{r}}+\lambda_{\mathrm{g}}} \right)  = 0.15$. Hence, the current sample data are unable to robustly constrain the dependence of the damping timescale upon wavelength.

\section{Discussions} \label{sec:discussion}
The observed damping timescales ($\tau_{\mathrm{obs}}$) in this work are more consistent with the theoretical predictions of \cite{Zhou2024} than that of \cite{Kelly2009}. Compared with previous observational studies, we carefully select sources to ensure that the damping timescales are not biased. In order to obtain the unbiased damping timescales, \textit{the theoretical/intrinsic damping timescale rather than the measured one should be smaller than the $10\%$ baseline.} We construct two well-defined samples in this work, and each source in our samples follows the condition that the theoretical damping timescale (i.e., $\tau_{\mathrm{th,Kelly+2009}}$ or $\tau_{\mathrm{th,Zhou+2024}}$) is smaller than the $10\%$ baseline in the rest frame, i.e., $\tau_{\mathrm{obs}}$ in this work are unbiased. Our measured damping timescales are inconsistent with the relationship of \cite{Kelly2009} (as shown in Section \ref{subsec:4.1}, and Fig.~\ref{Fig3}), possibly because this relationship is biased. In addition, \cite{Burke2021} also established a similar relationship between the damping timescale and the black hole mass. \cite{Zhou2024} has shown that the sample of \cite{Burke2021} is biased.

We obtained the dependencies of $\tau_{\mathrm{obs}}$ upon $L_{\mathrm{bol}}$ and $\lambda_{\mathrm{rest}}$ by our well-defined sample, i.e., $\log_{10}(\tau_{\mathrm{obs}}) = 0.72^{+0.18}_{-0.13} \log_{10}(L_{\mathrm{bol}}) + 1.19 \log_{10}(\lambda_{\mathrm{rest}}) - 34.2^{+5.7}_{-7.9}$ (as shown in Section \ref{sec:parameters fit}, Fig. \ref{Fig4} and \ref{Fig6}). There is a positive correlation between $\tau_{\mathrm{obs}}$ and $L_{\mathrm{bol}}$, and the coefficient ($a = 0.72^{+0.18}_{-0.13}$) of $L_{\mathrm{bol}}$ term is close to that of \cite{Zhou2024} ($a_{\mathrm{Zhou+2024}} = 0.65$), and the parameter $b = -34.2^{+5.7}_{-7.9}$ is also close to that of \cite{Zhou2024} ($b_{\mathrm{Zhou+2024}} = -30.9$). Hence, this suggests that the origin of AGN UV/optical variations can be described by the CHAR model; that is, the UV/optical variations of AGN are caused by the thermal fluctuations triggered by turbulence in the accretion disk, whose characteristic timescale depends mostly upon the mass accretion rate (or luminosity). In addition, as shown in Fig.~\ref{Fig4}, the CL-AGN NGC 5273 has different damping timescales in the low and high states. Therefore, our results suggest that the damping timescale might be used to measure the luminosity rather than the black hole mass of quasars. 

The unavoidable magnetohydrodynamic (MHD) turbulence plays an important role in the SMBH accretion process as it is responsible for removing the angular momentum of the accreted gas \citep[e.g.,][]{Balbus1991, Balbus2003}. The MHD turbulence in the accretion disk can drive significant variability in the UV/optical emission of the accretion disks. \cite{Zhou2024} demonstrated that in the MHD turbulence scenario, the UV/optical luminosity should vary on the thermal timescale, and the damping timescale is the average of thermal timescales at different radii of the accretion disk. Dwarf galaxies exhibit flux variations on short timescales because of their low luminosities. That is, the damping timescales of dwarf galaxies can be well constrained even for short-duration light curves. Hence, dwarf galaxies are a promising class of targets that can be used to study the MHD turbulence-driven thermal fluctuations in accretion disks. 

\section{Conclusion \label{sec:conclusion}}
We have constructed unbiased samples in which each source in our sample satisfies the criterion that the intrinsic/expected damping timescale is smaller than $10\%$ of the baseline. We have obtained the best-fit parameters between the damping timescales ($\tau_{\mathrm{obs}}$) and the bolometric luminosity ($L_{\mathrm{bol}}$) and the wavelength in the rest frame ($\lambda_{\mathrm{rest}}$), i.e., $\log_{10}(\tau_{\mathrm{obs}}) = 0.72^{+0.18}_{-0.13} \log_{10}(L_{\mathrm{bol}}) + 1.19 \log_{10}(\lambda_{\mathrm{rest}}) - 34.2^{+5.7}_{-7.9}$. Our results are roughly consistent with those of \cite{Zhou2024} which is based on the CHAR model \citep{Sun2020}. This agreement supports the idea that the observed UV/optical variability is due to thermal fluctuations in SMBH accretion disks. It is important to note that these samples are unable to robustly test the dependence of the damping timescale upon wavelength. In the future, sources with a larger wavelength range are needed to better test our conclusion and put strong constraints on the SMBH accretion physics.

\section{Acknowledgments}
G.W.R. would like to thank Dr.~Q.Z.~Yu for helpful discussions. We acknowledge support from the National Key R\&D Program of China (No. 2023YFA1607903). G.W.R., S.Y.Z., and M.Y.S. acknowledge support from the National Natural Science Foundation of China (NSFC-12322303) and the Natural Science Foundation of Fujian Province of China (No. 2022J06002). Y.Q.X. acknowledges support from the National Natural Science Foundation of China (NSFC-12025303). 

Based on observations obtained with the Samuel Oschin Telescope 48 inch and the 60 inch Telescope at the Palomar Observatory as part of the Zwicky Transient Facility project \citep{ZTF-DOI}. Z.T.F. is supported by the National Science Foundation under grants No. AST-1440341 and AST-2034437 and a collaboration including current partners Caltech, IPAC, the Weizmann Institute for Science, the Oskar Klein Center at Stockholm University, the University of Maryland, Deutsches Elektronen-Synchrotron and Humboldt University, the TANGO Consortium of Taiwan, the University of Wisconsin at Milwaukee, Trinity College Dublin, Lawrence Livermore National Laboratories, IN2P3, University of Warwick, Ruhr University Bochum, Northwestern University, and former partners the University of Washington, Los Alamos National Laboratories, and Lawrence Berkeley National Laboratories. Operations are conducted by COO, IPAC, and UW.
\\

\facilities{PO: 1.2m}

\software{Astropy \citep{Astropy2013}, emcee \citep{Foreman-Mackey2013}, Matplotlib \citep{Hunter2007}, Numpy \citep{Harris2020}, Scipy \citep{Virtanen2020}, taufit \citep{Burke2021}.}





\appendix

\renewcommand{\thefigure}{B\arabic{figure}}
\setcounter{figure}{0} 
\renewcommand{\thetable}{A\arabic{table}}

\section{Results of DRW fitting to the light curve samples in this work}
\setlength\tabcolsep{1pt}
\begin{longtable*}{cccccccccc}
\caption{Sample with $\tau_{\mathrm{th,Kelly+2009}}/\mathrm{baseline} < 10\% $}

\label{Tab1}\\

\hline\hline
Object & $z$ & $\log_{10}(M_{\mathrm{BH}})$ & $\log_{10}(L_{\mathrm{bol}})$ & band & $\lambda_{\mathrm{rest}}$ & $\log_{10}(\tau_{\mathrm{obs}})$ & $\log_{10}(\tau_{\mathrm{th,Kelly+2009}})$ & baseline & $\frac{\tau_{\mathrm{th,Kelly+2009}}}{\mathrm{baseline}}$  \\
& & $[M_{\odot}]$ & $[\mathrm{erg\, s^{-1} }]$ && $[\mathrm{\AA}]$ & [days] & [days] & [days] & \\
\hline
\endhead
\hline\hline
\multicolumn{10}{r}{\textit{Continued on next page}} \\ 
\endfoot
\hline\hline

\endlastfoot
1RXS J062334.6+644542  	&	0.086 	&	7.28 	&	44.92 	&	$zg$	&	4375	&	    2.33$^{+0.30}_{-0.16}$  	&	1.60 	&	1877	&	  2.12$\%$  \\
1RXS J062334.6+644542  	&	0.086 	&	7.28 	&	44.92 	&	$zr$	&	5862	&	    2.49$^{+0.43}_{-0.20}$  	&	1.60 	&	1868	&	  2.13$\%$  \\
1RXS J081749.0+025104  	&	0.106 	&	7.83 	&	45.39 	&	$zg$	&	4297	&	    2.52$^{+0.42}_{-0.22}$  	&	1.97 	&	1844	&	  5.06$\%$  \\
1RXS J081749.0+025104  	&	0.106 	&	7.83 	&	45.39 	&	$zr$	&	5758	&	    2.67$^{+0.51}_{-0.24}$  	&	1.97 	&	1833	&	  5.09$\%$  \\
1RXS J210932.5+353245  	&	0.201 	&	8.04 	&	45.94 	&	$zg$	&	3958	&	    2.76$^{+0.44}_{-0.21}$  	&	1.96 	&	1698	&	  5.37$\%$  \\
1RXS J214532.2+110255  	&	0.209 	&	8.17 	&	46.03 	&	$zg$	&	3933	&	    2.58$^{+0.49}_{-0.20}$  	&	2.05 	&	1654	&	  6.78$\%$  \\
1RXS J214532.2+110255  	&	0.209 	&	8.17 	&	46.03 	&	$zr$	&	5270	&	    2.50$^{+0.40}_{-0.16}$  	&	2.05 	&	1654	&	  6.78$\%$  \\
2MASS J06543417+0703210  	&	0.024 	&	6.74 	&	44.23 	&	$zg$	&	4642	&	    2.45$^{+0.44}_{-0.23}$  	&	1.34 	&	1993	&	  1.10$\%$  \\
2MASS J06543417+0703210  	&	0.024 	&	6.74 	&	44.23 	&	$zr$	&	6220	&	    2.26$^{+0.32}_{-0.19}$  	&	1.34 	&	1982	&	  1.10$\%$  \\
2MASS J17465953+6836303  	&	0.064 	&	7.43 	&	44.54 	&	$zg$	&	4468	&	    2.25$^{+0.35}_{-0.16}$  	&	1.92 	&	1918	&	  4.34$\%$  \\
2MASS J17465953+6836303  	&	0.064 	&	7.43 	&	44.54 	&	$zr$	&	5987	&	    2.21$^{+0.30}_{-0.15}$  	&	1.92 	&	1923	&	  4.33$\%$  \\
2MASX J17311341+1442561  	&	0.082 	&	7.56 	&	45.11 	&	$zg$	&	4392	&	    2.63$^{+0.46}_{-0.24}$  	&	1.81 	&	1884	&	  3.43$\%$  \\
2MASX J20350566+2603301  	&	0.049 	&	7.18 	&	44.63 	&	$zg$	&	4533	&	    1.77$^{+0.15}_{-0.13}$  	&	1.62 	&	1918	&	  2.17$\%$  \\
2MASX J20350566+2603301  	&	0.049 	&	7.18 	&	44.63 	&	$zr$	&	6074	&	    2.00$^{+0.22}_{-0.15}$  	&	1.62 	&	1936	&	  2.15$\%$  \\
2MASX J21192912+3332566  	&	0.051 	&	7.63 	&	44.73 	&	$zg$	&	4523	&	    1.75$^{+0.21}_{-0.14}$  	&	2.04 	&	1900	&	  5.77$\%$  \\
2MASX J21344509-2725557  	&	0.067 	&	6.94 	&	45.11 	&	$zg$	&	4455	&	    2.47$^{+0.51}_{-0.26}$  	&	1.17 	&	1842	&	  0.80$\%$  \\
2MASX J21355399+4728217  	&	0.026 	&	7.40 	&	44.37 	&	$zg$	&	4634	&	    2.30$^{+0.36}_{-0.19}$  	&	1.96 	&	1985	&	  4.59$\%$  \\
2MASX J21355399+4728217  	&	0.026 	&	7.40 	&	44.37 	&	$zr$	&	6209	&	    1.52$^{+0.11}_{-0.09}$  	&	1.96 	&	1976	&	  4.62$\%$  \\
2MASX J23075724+4016393  	&	0.073 	&	7.61 	&	45.00 	&	$zg$	&	4431	&	    3.10$^{+0.53}_{-0.31}$  	&	1.91 	&	1863	&	  4.36$\%$  \\
3C 382  	&	0.058 	&	8.01 	&	45.62 	&	$zg$	&	4493	&	    2.78$^{+0.58}_{-0.28}$  	&	2.06 	&	1927	&	  5.96$\%$  \\
3C 382  	&	0.058 	&	8.01 	&	45.62 	&	$zr$	&	6020	&	    2.60$^{+0.55}_{-0.25}$  	&	2.06 	&	1936	&	  5.93$\%$  \\
Arp 151  	&	0.021 	&	6.67 	&	44.11 	&	$zg$	&	4657	&	    1.99$^{+0.19}_{-0.13}$  	&	1.32 	&	1859	&	  1.12$\%$  \\
Arp 151  	&	0.021 	&	6.67 	&	44.11 	&	$zr$	&	6240	&	    2.22$^{+0.31}_{-0.17}$  	&	1.32 	&	1734	&	  1.20$\%$  \\
CTS 103  	&	0.012 	&	6.11 	&	43.63 	&	$zg$	&	4696	&	    1.78$^{+0.26}_{-0.21}$  	&	0.94 	&	1894	&	  0.46$\%$  \\
CTS 103  	&	0.012 	&	6.11 	&	43.63 	&	$zr$	&	6293	&	    2.15$^{+0.34}_{-0.22}$  	&	0.94 	&	1904	&	  0.46$\%$  \\
ESO 424-12  	&	0.010 	&	6.07 	&	43.48 	&	$zg$	&	4707	&	    0.94$^{+0.12}_{-0.11}$  	&	0.96 	&	2022	&	  0.45$\%$  \\
ESO 490-26  	&	0.025 	&	7.15 	&	44.57 	&	$zg$	&	4636	&	    2.06$^{+0.32}_{-0.22}$  	&	1.62 	&	1779	&	  2.34$\%$  \\
ESO 511-30  	&	0.023 	&	7.23 	&	44.52 	&	$zg$	&	4646	&	    1.96$^{+0.29}_{-0.19}$  	&	1.72 	&	1869	&	  2.81$\%$  \\
ESO 511-30  	&	0.023 	&	7.23 	&	44.52 	&	$zr$	&	6226	&	    2.19$^{+0.40}_{-0.24}$  	&	1.72 	&	1868	&	  2.81$\%$  \\
ESO 578-9  	&	0.035 	&	7.61 	&	44.54 	&	$zg$	&	4592	&	    1.79$^{+0.18}_{-0.14}$  	&	2.10 	&	1852	&	  6.80$\%$  \\
ESO 578-9  	&	0.035 	&	7.61 	&	44.54 	&	$zr$	&	6154	&	    1.86$^{+0.20}_{-0.16}$  	&	2.10 	&	1862	&	  6.76$\%$  \\
H 2106-099  	&	0.027 	&	7.51 	&	44.35 	&	$zg$	&	4629	&	    1.93$^{+0.18}_{-0.13}$  	&	2.08 	&	1930	&	  6.23$\%$  \\
H 2106-099  	&	0.027 	&	7.51 	&	44.35 	&	$zr$	&	6203	&	    1.90$^{+0.17}_{-0.12}$  	&	2.08 	&	1938	&	  6.20$\%$  \\
HE 0309-2057  	&	0.067 	&	7.79 	&	45.21 	&	$zg$	&	4456	&	    2.57$^{+0.48}_{-0.25}$  	&	2.01 	&	1780	&	  5.75$\%$  \\
HE 0309-2057  	&	0.067 	&	7.79 	&	45.21 	&	$zr$	&	5971	&	    2.54$^{+0.50}_{-0.26}$  	&	2.01 	&	1779	&	  5.75$\%$  \\
HE 1143-1810  	&	0.033 	&	7.39 	&	44.95 	&	$zg$	&	4603	&	    2.29$^{+0.38}_{-0.21}$  	&	1.70 	&	1840	&	  2.72$\%$  \\
HE 1143-1810  	&	0.033 	&	7.39 	&	44.95 	&	$zr$	&	6168	&	    2.13$^{+0.29}_{-0.18}$  	&	1.70 	&	1826	&	  2.74$\%$  \\
HE 1310-1051  	&	0.035 	&	7.49 	&	44.46 	&	$zg$	&	4594	&	    1.71$^{+0.16}_{-0.12}$  	&	2.01 	&	1852	&	  5.53$\%$  \\
HE 1310-1051  	&	0.035 	&	7.49 	&	44.46 	&	$zr$	&	6156	&	    1.77$^{+0.19}_{-0.15}$  	&	2.01 	&	1843	&	  5.55$\%$  \\
HS 0328+0528  	&	0.045 	&	7.40 	&	44.58 	&	$zg$	&	4549	&	    2.14$^{+0.24}_{-0.15}$  	&	1.87 	&	1951	&	  3.80$\%$  \\
HS 0328+0528  	&	0.045 	&	7.40 	&	44.58 	&	$zr$	&	6096	&	    2.20$^{+0.29}_{-0.17}$  	&	1.87 	&	1838	&	  4.03$\%$  \\
IC 2921  	&	0.044 	&	7.70 	&	44.67 	&	$zr$	&	6102	&	    1.11$^{+0.09}_{-0.09}$  	&	2.14 	&	1946	&	  7.09$\%$  \\
IRAS 05078+1626  	&	0.017 	&	6.88 	&	44.64 	&	$zg$	&	4672	&	    2.41$^{+0.47}_{-0.22}$  	&	1.31 	&	2008	&	  1.02$\%$  \\
IRAS 05078+1626  	&	0.017 	&	6.88 	&	44.64 	&	$zr$	&	6260	&	    2.73$^{+0.61}_{-0.32}$  	&	1.31 	&	1996	&	  1.02$\%$  \\
IRAS 05320+4009  	&	0.020 	&	6.59 	&	43.99 	&	$zg$	&	4660	&	    1.46$^{+0.30}_{-0.22}$  	&	1.28 	&	1018	&	  1.87$\%$  \\
IRAS 05320+4009  	&	0.020 	&	6.59 	&	43.99 	&	$zr$	&	6244	&	    1.12$^{+0.19}_{-0.16}$  	&	1.28 	&	1035	&	  1.84$\%$  \\
IRAS 15091-2107  	&	0.044 	&	6.94 	&	44.98 	&	$zg$	&	4551	&	    2.40$^{+0.43}_{-0.22}$  	&	1.23 	&	1857	&	  0.91$\%$  \\
LEDA 136513  	&	0.042 	&	7.80 	&	44.66 	&	$zg$	&	4559	&	    1.85$^{+0.16}_{-0.12}$  	&	2.25 	&	1958	&	  9.08$\%$  \\
LEDA 136513  	&	0.042 	&	7.80 	&	44.66 	&	$zr$	&	6110	&	    1.68$^{+0.23}_{-0.18}$  	&	2.25 	&	1948	&	  9.13$\%$  \\
LEDA 138501  	&	0.050 	&	8.01 	&	45.32 	&	$zg$	&	4528	&	    3.04$^{+0.54}_{-0.32}$  	&	2.19 	&	1883	&	  8.23$\%$  \\
LEDA 168563  	&	0.028 	&	7.88 	&	44.88 	&	$zg$	&	4622	&	    2.56$^{+0.52}_{-0.27}$  	&	2.24 	&	1975	&	  8.80$\%$  \\
LEDA 168563  	&	0.028 	&	7.88 	&	44.88 	&	$zr$	&	6194	&	    2.50$^{+0.53}_{-0.26}$  	&	2.24 	&	1975	&	  8.80$\%$  \\
LEDA 36114  	&	0.061 	&	7.85 	&	45.11 	&	$zg$	&	4479	&	    2.48$^{+0.46}_{-0.21}$  	&	2.11 	&	1790	&	  7.20$\%$  \\
LEDA 36114  	&	0.061 	&	7.85 	&	45.11 	&	$zr$	&	6002	&	    2.33$^{+0.35}_{-0.18}$  	&	2.11 	&	1793	&	  7.18$\%$  \\
LEDA 70195  	&	0.034 	&	6.87 	&	44.65 	&	$zg$	&	4598	&	    1.49$^{+0.22}_{-0.15}$  	&	1.29 	&	548	&	  3.56$\%$  \\
LEDA 70195  	&	0.034 	&	6.87 	&	44.65 	&	$zr$	&	6161	&	    1.69$^{+0.33}_{-0.20}$  	&	1.29 	&	535	&	  3.64$\%$  \\
LEDA 75258  	&	0.016 	&	7.02 	&	43.74 	&	$zg$	&	4679	&	    1.79$^{+0.15}_{-0.12}$  	&	1.83 	&	2011	&	  3.36$\%$  \\
LEDA 75258  	&	0.016 	&	7.02 	&	43.74 	&	$zr$	&	6269	&	    1.94$^{+0.19}_{-0.13}$  	&	1.83 	&	1877	&	  3.60$\%$  \\
LEDA 86073  	&	0.024 	&	7.51 	&	44.27 	&	$zg$	&	4642	&	    2.48$^{+0.55}_{-0.28}$  	&	2.11 	&	1797	&	  7.17$\%$  \\
LEDA 86073  	&	0.024 	&	7.51 	&	44.27 	&	$zr$	&	6220	&	    1.09$^{+0.12}_{-0.11}$  	&	2.11 	&	1578	&	  8.16$\%$  \\
LEDA 90334  	&	0.010 	&	6.61 	&	43.61 	&	$zg$	&	4704	&	    1.30$^{+0.09}_{-0.08}$  	&	1.46 	&	2018	&	  1.43$\%$  \\
LEDA 90334  	&	0.010 	&	6.61 	&	43.61 	&	$zr$	&	6304	&	    1.32$^{+0.09}_{-0.08}$  	&	1.46 	&	2006	&	  1.44$\%$  \\
MCG+9-21-96  	&	0.030 	&	7.67 	&	44.66 	&	$zg$	&	4614	&	    1.44  $^{+  0.14}_{-0.11}$  	&	2.11 	&	1903	&	  6.77$\%$  \\
MCG-2-14-9  	&	0.028 	&	7.02 	&	44.27 	&	$zg$	&	4623	&	    2.41$^{+0.42}_{-0.22}$  	&	1.61 	&	1987	&	  2.05$\%$  \\
MCG-2-14-9  	&	0.028 	&	7.02 	&	44.27 	&	$zr$	&	6195	&	    2.31$^{+0.38}_{-0.20}$  	&	1.61 	&	1836	&	  2.22$\%$  \\
MCG-3-4-72  	&	0.043 	&	7.03 	&	44.76 	&	$zg$	&	4557	&	    2.95$^{+0.60}_{-0.37}$  	&	1.41 	&	1846	&	  1.39$\%$  \\
MCG-3-4-72  	&	0.043 	&	7.03 	&	44.76 	&	$zr$	&	6107	&	    2.94$^{+0.59}_{-0.37}$  	&	1.41 	&	1852	&	  1.39$\%$  \\
Mrk 10  	&	0.029 	&	7.19 	&	44.29 	&	$zg$	&	4620	&	    1.89$^{+0.18}_{-0.14}$  	&	1.78 	&	1985	&	  3.04$\%$  \\
Mrk 10  	&	0.029 	&	7.19 	&	44.29 	&	$zr$	&	6190	&	    1.81$^{+0.16}_{-0.13}$  	&	1.78 	&	1973	&	  3.05$\%$  \\
Mrk 1018  	&	0.043 	&	7.81 	&	44.96 	&	$zg$	&	4559	&	    1.91$^{+0.22}_{-0.17}$  	&	2.13 	&	1861	&	  7.25$\%$  \\
Mrk 1018  	&	0.043 	&	7.81 	&	44.96 	&	$zr$	&	6109	&	    1.68$^{+0.20}_{-0.15}$  	&	2.13 	&	1855	&	  7.27$\%$  \\
Mrk 1044  	&	0.017 	&	6.45 	&	43.72 	&	$zg$	&	4672	&	    1.94$^{+0.20}_{-0.16}$  	&	1.25 	&	1885	&	  0.94$\%$  \\
Mrk 1044  	&	0.017 	&	6.45 	&	43.72 	&	$zr$	&	6261	&	    1.99$^{+0.23}_{-0.16}$  	&	1.25 	&	1891	&	  0.94$\%$  \\
Mrk 110  	&	0.035 	&	7.29 	&	45.06 	&	$zg$	&	4591	&	    3.11$^{+0.57}_{-0.39}$  	&	1.55 	&	1975	&	  1.80$\%$  \\
Mrk 110  	&	0.035 	&	7.29 	&	45.06 	&	$zr$	&	6152	&	    3.03$^{+0.58}_{-0.36}$  	&	1.55 	&	1961	&	  1.81$\%$  \\
Mrk 1148  	&	0.064 	&	7.75 	&	45.30 	&	$zr$	&	5986	&	    2.50$^{+0.54}_{-0.29}$  	&	1.93 	&	937	&	  9.08$\%$  \\
Mrk 1310  	&	0.019 	&	6.21 	&	43.82 	&	$zg$	&	4662	&	    1.83$^{+0.22}_{-0.16}$  	&	0.96 	&	1865	&	  0.49$\%$  \\
Mrk 1498  	&	0.056 	&	7.19 	&	45.38 	&	$zg$	&	4502	&	    1.10$^{+0.07}_{-0.06}$  	&	1.32 	&	1922	&	  1.09$\%$  \\
Mrk 1498  	&	0.056 	&	7.19 	&	45.38 	&	$zr$	&	6032	&	    1.97$^{+0.18}_{-0.12}$  	&	1.32 	&	1919	&	  1.09$\%$  \\
Mrk 290  	&	0.030 	&	7.28 	&	44.51 	&	$zg$	&	4613	&	    2.21$^{+0.26}_{-0.15}$  	&	1.78 	&	1962	&	  3.07$\%$  \\
Mrk 290  	&	0.030 	&	7.28 	&	44.51 	&	$zr$	&	6181	&	    2.24$^{+0.26}_{-0.16}$  	&	1.78 	&	1965	&	  3.07$\%$  \\
Mrk 359  	&	0.017 	&	6.05 	&	43.75 	&	$zg$	&	4675	&	    1.37$^{+0.11}_{-0.10}$  	&	0.83 	&	1924	&	  0.35$\%$  \\
Mrk 359  	&	0.017 	&	6.05 	&	43.75 	&	$zr$	&	6264	&	    1.11$^{+0.10}_{-0.09}$  	&	0.83 	&	1918	&	  0.35$\%$  \\
Mrk 50  	&	0.024 	&	7.42 	&	44.31 	&	$zg$	&	4643	&	    2.22$^{+0.31}_{-0.18}$  	&	2.00 	&	1878	&	  5.32$\%$  \\
Mrk 50  	&	0.024 	&	7.42 	&	44.31 	&	$zr$	&	6222	&	    2.46$^{+0.50}_{-0.23}$  	&	2.00 	&	1865	&	  5.36$\%$  \\
Mrk 507  	&	0.055 	&	6.08 	&	44.29 	&	$zg$	&	4505	&	    2.42$^{+0.62}_{-0.31}$  	&	0.63 	&	1541	&	  0.28$\%$  \\
Mrk 509  	&	0.035 	&	8.05 	&	45.27 	&	$zg$	&	4594	&	    3.10$^{+0.59}_{-0.38}$  	&	2.25 	&	1924	&	  9.24$\%$  \\
Mrk 595  	&	0.027 	&	7.41 	&	44.10 	&	$zg$	&	4627	&	    2.11$^{+0.30}_{-0.18}$  	&	2.08 	&	1888	&	  6.37$\%$  \\
Mrk 595  	&	0.027 	&	7.41 	&	44.10 	&	$zr$	&	6200	&	    1.77$^{+0.19}_{-0.15}$  	&	2.08 	&	1879	&	  6.40$\%$  \\
Mrk 618  	&	0.035 	&	7.72 	&	44.51 	&	$zg$	&	4594	&	    2.39$^{+0.36}_{-0.21}$  	&	2.23 	&	1972	&	  8.61$\%$  \\
Mrk 618  	&	0.035 	&	7.72 	&	44.51 	&	$zr$	&	6156	&	    2.30$^{+0.40}_{-0.20}$  	&	2.23 	&	1845	&	  9.20$\%$  \\
Mrk 684  	&	0.046 	&	6.78 	&	44.46 	&	$zg$	&	4544	&	    2.32$^{+0.39}_{-0.19}$  	&	1.28 	&	1886	&	  1.01$\%$  \\
Mrk 684  	&	0.046 	&	6.78 	&	44.46 	&	$zr$	&	6089	&	    2.07$^{+0.22}_{-0.13}$  	&	1.28 	&	1888	&	  1.01$\%$  \\
Mrk 705  	&	0.029 	&	7.08 	&	44.31 	&	$zg$	&	4621	&	    2.00$^{+0.21}_{-0.15}$  	&	1.65 	&	1988	&	  2.25$\%$  \\
Mrk 705  	&	0.029 	&	7.08 	&	44.31 	&	$zr$	&	6192	&	    2.17$^{+0.26}_{-0.16}$  	&	1.65 	&	1973	&	  2.26$\%$  \\
Mrk 728  	&	0.036 	&	6.66 	&	44.41 	&	$zr$	&	6149	&	    1.04$^{+0.11}_{-0.10}$  	&	1.18 	&	1957	&	  0.77$\%$  \\
Mrk 732  	&	0.029 	&	6.59 	&	44.17 	&	$zr$	&	6188	&	    1.81$^{+0.20}_{-0.17}$  	&	1.21 	&	1969	&	  0.82$\%$  \\
Mrk 739E  	&	0.029 	&	6.99 	&	44.26 	&	$zg$	&	4617	&	    2.17$^{+0.29}_{-0.17}$  	&	1.58 	&	1861	&	  2.04$\%$  \\
Mrk 739E  	&	0.029 	&	6.99 	&	44.26 	&	$zr$	&	6187	&	    1.91$^{+0.18}_{-0.12}$  	&	1.58 	&	1972	&	  1.93$\%$  \\
Mrk 771  	&	0.064 	&	7.76 	&	44.85 	&	$zg$	&	4468	&	    2.65$^{+0.51}_{-0.25}$  	&	2.13 	&	1785	&	  7.56$\%$  \\
Mrk 783  	&	0.067 	&	7.30 	&	45.11 	&	$zg$	&	4455	&	    1.63$^{+0.11}_{-0.09}$  	&	1.54 	&	1801	&	  1.93$\%$  \\
Mrk 783  	&	0.067 	&	7.30 	&	45.11 	&	$zr$	&	5969	&	    2.00$^{+0.19}_{-0.14}$  	&	1.54 	&	1785	&	  1.94$\%$  \\
Mrk 79  	&	0.022 	&	7.61 	&	44.54 	&	$zg$	&	4650	&	    2.12$^{+0.22}_{-0.14}$  	&	2.10 	&	1999	&	  6.30$\%$  \\
Mrk 79  	&	0.022 	&	7.61 	&	44.54 	&	$zr$	&	6231	&	    2.25$^{+0.27}_{-0.16}$  	&	2.10 	&	1980	&	  6.36$\%$  \\
Mrk 817  	&	0.031 	&	7.59 	&	44.60 	&	$zg$	&	4609	&	    2.34$^{+0.31}_{-0.18}$  	&	2.06 	&	1932	&	  5.94$\%$  \\
Mrk 817  	&	0.031 	&	7.59 	&	44.60 	&	$zr$	&	6176	&	    2.37$^{+0.36}_{-0.18}$  	&	2.06 	&	1936	&	  5.93$\%$  \\
Mrk 9  	&	0.040 	&	7.60 	&	44.34 	&	$zg$	&	4571	&	    1.93$^{+0.16}_{-0.11}$  	&	2.18 	&	1964	&	  7.71$\%$  \\
Mrk 9  	&	0.040 	&	7.60 	&	44.34 	&	$zr$	&	6125	&	    2.17$^{+0.26}_{-0.15}$  	&	2.18 	&	1952	&	  7.75$\%$  \\
Mrk 975  	&	0.050 	&	7.45 	&	44.83 	&	$zg$	&	4527	&	    2.08$^{+0.25}_{-0.15}$  	&	1.82 	&	1855	&	  3.56$\%$  \\
NGC 3080  	&	0.035 	&	6.84 	&	44.01 	&	$zg$	&	4590	&	    2.19$^{+0.30}_{-0.18}$  	&	1.53 	&	1975	&	  1.72$\%$  \\
NGC 3080  	&	0.035 	&	6.84 	&	44.01 	&	$zr$	&	6151	&	    2.00$^{+0.21}_{-0.14}$  	&	1.53 	&	1960	&	  1.73$\%$  \\
NGC 3227  	&	0.003 	&	6.77 	&	43.67 	&	$zg$	&	4737	&	    1.71$^{+0.15}_{-0.13}$  	&	1.60 	&	2036	&	  1.96$\%$  \\
NGC 3227  	&	0.003 	&	6.77 	&	43.67 	&	$zr$	&	6348	&	    1.86$^{+0.16}_{-0.13}$  	&	1.60 	&	2021	&	  1.97$\%$  \\
NGC 3822  	&	0.019 	&	7.43 	&	43.91 	&	$zg$	&	4662	&	    1.59$^{+0.18}_{-0.15}$  	&	2.18 	&	1877	&	  8.06$\%$  \\
NGC 3822  	&	0.019 	&	7.43 	&	43.91 	&	$zr$	&	6247	&	    1.03$^{+0.16}_{-0.13}$  	&	2.18 	&	1992	&	  7.60$\%$  \\
NGC 4051  	&	0.002 	&	6.13 	&	42.59 	&	$zg$	&	4743	&	    1.58$^{+0.11}_{-0.10}$  	&	1.40 	&	1956	&	  1.28$\%$  \\
NGC 4051  	&	0.002 	&	6.13 	&	42.59 	&	$zr$	&	6356	&	    1.65$^{+0.13}_{-0.11}$  	&	1.40 	&	1884	&	  1.33$\%$  \\
NGC 4253  	&	0.013 	&	6.82 	&	43.75 	&	$zg$	&	4692	&	    2.12$^{+0.25}_{-0.16}$  	&	1.62 	&	1902	&	  2.19$\%$  \\
NGC 4253  	&	0.013 	&	6.82 	&	43.75 	&	$zr$	&	6288	&	    2.20$^{+0.26}_{-0.16}$  	&	1.62 	&	2006	&	  2.08$\%$  \\
NGC 4395  	&	0.001 	&	5.45 	&	41.68 	&	$zg$	&	4748	&	    0.95$^{+0.07}_{-0.07}$  	&	1.08 	&	1924	&	  0.62$\%$  \\
NGC 4395  	&	0.001 	&	5.45 	&	41.68 	&	$zr$	&	6362	&	    0.99$^{+0.07}_{-0.07}$  	&	1.08 	&	2024	&	  0.59$\%$  \\
NGC 4593  	&	0.008 	&	6.88 	&	44.00 	&	$zg$	&	4714	&	    2.34$^{+0.48}_{-0.30}$  	&	1.58 	&	1900	&	  2.00$\%$  \\
NGC 4593  	&	0.008 	&	6.88 	&	44.00 	&	$zr$	&	6316	&	    1.64$^{+0.19}_{-0.15}$  	&	1.58 	&	1893	&	  2.01$\%$  \\
NGC 4748  	&	0.014 	&	6.41 	&	43.63 	&	$zg$	&	4687	&	    2.06$^{+0.27}_{-0.18}$  	&	1.25 	&	1886	&	  0.94$\%$  \\
NGC 4748  	&	0.014 	&	6.41 	&	43.63 	&	$zr$	&	6280	&	    2.29$^{+0.40}_{-0.22}$  	&	1.25 	&	1883	&	  0.94$\%$  \\
NGC 6814  	&	0.006 	&	7.04 	&	43.52 	&	$zr$	&	6332	&	    1.10$^{+0.19}_{-0.16}$  	&	1.94 	&	933	&	  9.34$\%$  \\
NGC 7214  	&	0.023 	&	7.44 	&	43.99 	&	$zg$	&	4648	&	    1.95$^{+0.23}_{-0.17}$  	&	2.16 	&	1916	&	  7.54$\%$  \\
NGC 7214  	&	0.023 	&	7.44 	&	43.99 	&	$zr$	&	6228	&	    2.29$^{+0.36}_{-0.21}$  	&	2.16 	&	1913	&	  7.56$\%$  \\
NGC 7811  	&	0.025 	&	6.70 	&	44.01 	&	$zg$	&	4635	&	    1.96$^{+0.20}_{-0.14}$  	&	1.39 	&	1926	&	  1.27$\%$  \\
NGC 7811  	&	0.025 	&	6.70 	&	44.01 	&	$zr$	&	6211	&	    2.11$^{+0.28}_{-0.17}$  	&	1.39 	&	1906	&	  1.29$\%$  \\
PG 1149-110  	&	0.049 	&	7.68 	&	44.76 	&	$zg$	&	4533	&	    2.21$^{+0.32}_{-0.19}$  	&	2.08 	&	1814	&	  6.63$\%$  \\
PG 1149-110  	&	0.049 	&	7.68 	&	44.76 	&	$zr$	&	6074	&	    1.78$^{+0.17}_{-0.12}$  	&	2.08 	&	1802	&	  6.67$\%$  \\
PG 1202+282  	&	0.166 	&	8.10 	&	45.73 	&	$zg$	&	4078	&	    2.91$^{+0.53}_{-0.28}$  	&	2.11 	&	1642	&	  7.85$\%$  \\
RBS 149  	&	0.067 	&	7.88 	&	44.96 	&	$zg$	&	4456	&	    2.23$^{+0.30}_{-0.17}$  	&	2.20 	&	1824	&	  8.69$\%$  \\
RBS 149  	&	0.067 	&	7.88 	&	44.96 	&	$zr$	&	5970	&	    2.24$^{+0.32}_{-0.17}$  	&	2.20 	&	1812	&	  8.75$\%$  \\
RBS 1645  	&	0.040 	&	7.24 	&	44.65 	&	$zg$	&	4570	&	    1.81$^{+0.14}_{-0.10}$  	&	1.68 	&	1958	&	  2.44$\%$  \\
RBS 1645  	&	0.040 	&	7.24 	&	44.65 	&	$zr$	&	6124	&	    1.75$^{+0.13}_{-0.10}$  	&	1.68 	&	1963	&	  2.44$\%$  \\
RHS 10  	&	0.110 	&	8.04 	&	45.47 	&	$zg$	&	4281	&	    3.08$^{+0.49}_{-0.29}$  	&	2.16 	&	1762	&	  8.20$\%$  \\
RHS 10  	&	0.110 	&	8.04 	&	45.47 	&	$zr$	&	5737	&	    2.93$^{+0.52}_{-0.28}$  	&	2.16 	&	1765	&	  8.19$\%$  \\
RHS 25  	&	0.043 	&	6.78 	&	44.52 	&	$zr$	&	6105	&	    2.68$^{+0.59}_{-0.42}$  	&	1.26 	&	730	&	  2.49$\%$  \\
RXJ 1313.8+3653  	&	0.067 	&	7.74 	&	44.79 	&	$zg$	&	4455	&	    1.71$^{+0.12}_{-0.10}$  	&	2.13 	&	1826	&	  7.39$\%$  \\
RXJ 1313.8+3653  	&	0.067 	&	7.74 	&	44.79 	&	$zr$	&	5969	&	    1.72$^{+0.12}_{-0.09}$  	&	2.13 	&	1823	&	  7.40$\%$  \\
SDSS J080327.38+084152.2  	&	0.047 	&	7.70 	&	44.80 	&	$zg$	&	4540	&	    2.36$^{+0.43}_{-0.22}$  	&	2.09 	&	1949	&	  6.31$\%$  \\
SDSS J080327.38+084152.2  	&	0.047 	&	7.70 	&	44.80 	&	$zr$	&	6084	&	    2.65$^{+0.58}_{-0.28}$  	&	2.09 	&	1938	&	  6.35$\%$  \\
SDSS J090436.96+553602.7  	&	0.037 	&	7.72 	&	44.54 	&	$zg$	&	4583	&	    1.81$^{+0.14}_{-0.10}$  	&	2.22 	&	1971	&	  8.42$\%$  \\
SDSS J090436.96+553602.7  	&	0.037 	&	7.72 	&	44.54 	&	$zr$	&	6141	&	    1.88$^{+0.16}_{-0.11}$  	&	2.22 	&	1957	&	  8.48$\%$  \\
SDSS J093527.09+261709.6  	&	0.122 	&	7.95 	&	45.55 	&	$zg$	&	4236	&	    3.21$^{+0.53}_{-0.36}$  	&	2.03 	&	1822	&	  5.88$\%$  \\
SDSS J093527.09+261709.6  	&	0.122 	&	7.95 	&	45.55 	&	$zr$	&	5676	&	    3.13$^{+0.54}_{-0.35}$  	&	2.03 	&	1809	&	  5.92$\%$  \\
SDSS J134628.41+192243.2  	&	0.084 	&	7.95 	&	45.16 	&	$zr$	&	5874	&	    1.78$^{+0.17}_{-0.12}$  	&	2.19 	&	1797	&	  8.62$\%$  \\
SDSS J225051.71-085456.8  	&	0.065 	&	7.35 	&	44.88 	&	$zg$	&	4463	&	    1.40$^{+0.13}_{-0.10}$  	&	1.69 	&	1847	&	  2.65$\%$  \\
SDSS J225051.71-085456.8  	&	0.065 	&	7.35 	&	44.88 	&	$zr$	&	5980	&	    1.70$^{+0.15}_{-0.12}$  	&	1.69 	&	1856	&	  2.64$\%$  \\
UGC 12138  	&	0.025 	&	7.08 	&	44.21 	&	$zg$	&	4637	&	    2.44$^{+0.40}_{-0.21}$  	&	1.70 	&	1945	&	  2.58$\%$  \\
UGC 12138  	&	0.025 	&	7.08 	&	44.21 	&	$zr$	&	6214	&	    2.44$^{+0.41}_{-0.21}$  	&	1.70 	&	1943	&	  2.58$\%$  \\
UGC 3374  	&	0.020 	&	6.61 	&	44.93 	&	$zg$	&	4659	&	    2.48$^{+0.52}_{-0.27}$  	&	0.91 	&	1100	&	  0.74$\%$  \\
UGC 3374  	&	0.020 	&	6.61 	&	44.93 	&	$zr$	&	6243	&	    2.59$^{+0.53}_{-0.28}$  	&	0.91 	&	1779	&	  0.46$\%$  \\
UGC 3478  	&	0.013 	&	6.06 	&	43.40 	&	$zg$	&	4693	&	    2.41$^{+0.53}_{-0.26}$  	&	0.99 	&	2014	&	  0.49$\%$  \\
UGC 3601  	&	0.017 	&	7.33 	&	43.99 	&	$zg$	&	4673	&	    1.47$^{+0.11}_{-0.09}$  	&	2.05 	&	2008	&	  5.59$\%$  \\
UGC 3601  	&	0.017 	&	7.33 	&	43.99 	&	$zr$	&	6261	&	    1.65$^{+0.13}_{-0.10}$  	&	2.05 	&	1999	&	  5.61$\%$  \\
UGC 524  	&	0.036 	&	7.62 	&	44.30 	&	$zg$	&	4588	&	    2.08$^{+0.24}_{-0.15}$  	&	2.21 	&	1908	&	  8.50$\%$  \\
UGC 6728  	&	0.006 	&	5.55 	&	43.16 	&	$zg$	&	4725	&	    1.12$^{+0.08}_{-0.08}$  	&	0.56 	&	1890	&	  0.19$\%$  \\
UGC 6728  	&	0.006 	&	5.55 	&	43.16 	&	$zr$	&	6331	&	    1.19$^{+0.09}_{-0.08}$  	&	0.56 	&	1895	&	  0.19$\%$  \\
UM 614  	&	0.033 	&	7.16 	&	44.46 	&	$zg$	&	4602	&	    1.47$^{+0.11}_{-0.09}$  	&	1.67 	&	1876	&	  2.49$\%$  \\
UM 614  	&	0.033 	&	7.16 	&	44.46 	&	$zr$	&	6167	&	    1.68$^{+0.15}_{-0.11}$  	&	1.67 	&	1882	&	  2.49$\%$  \\
Was 49b  	&	0.063 	&	7.25 	&	45.05 	&	$zg$	&	4470	&	    0.89$^{+0.07}_{-0.07}$  	&	1.52 	&	1806	&	  1.83$\%$  \\
Was 49b  	&	0.063 	&	7.25 	&	45.05 	&	$zr$	&	5990	&	    0.80$^{+0.08}_{-0.07}$  	&	1.52 	&	1911	&	  1.73$\%$  \\
Z 121-75  	&	0.033 	&	7.27 	&	44.84 	&	$zg$	&	4601	&	    2.84$^{+0.62}_{-0.33}$  	&	1.63 	&	1979	&	  2.16$\%$  \\
Z 121-75  	&	0.033 	&	7.27 	&	44.84 	&	$zr$	&	6165	&	    2.95$^{+0.61}_{-0.37}$  	&	1.63 	&	1965	&	  2.17$\%$  \\
Z 122-55  	&	0.021 	&	7.13 	&	43.93 	&	$zg$	&	4654	&	    2.30$^{+0.38}_{-0.21}$  	&	1.86 	&	2002	&	  3.62$\%$  \\
Z 122-55  	&	0.021 	&	7.13 	&	43.93 	&	$zr$	&	6236	&	    1.77$^{+0.14}_{-0.11}$  	&	1.86 	&	1987	&	  3.65$\%$  \\
Z 215-46  	&	0.023 	&	7.50 	&	44.22 	&	$zg$	&	4646	&	    2.63$^{+0.56}_{-0.27}$  	&	2.12 	&	1879	&	  7.02$\%$  \\
Z 215-46  	&	0.023 	&	7.50 	&	44.22 	&	$zr$	&	6226	&	    2.88$^{+0.61}_{-0.35}$  	&	2.12 	&	1722	&	  7.66$\%$  \\
Z 31-72  	&	0.034 	&	5.94 	&	44.62 	&	$zr$	&	6158	&	    1.56$^{+0.21}_{-0.15}$  	&	0.35 	&	1537	&	  0.15$\%$  \\
Z 493-2  	&	0.025 	&	7.44 	&	44.20 	&	$zg$	&	4638	&	    1.84$^{+0.15}_{-0.11}$  	&	2.07 	&	1966	&	  5.98$\%$  \\
Z 493-2  	&	0.025 	&	7.44 	&	44.20 	&	$zr$	&	6216	&	    1.67$^{+0.13}_{-0.10}$  	&	2.07 	&	1980	&	  5.93$\%$  \\
Z 535-12  	&	0.048 	&	7.36 	&	44.75 	&	$zg$	&	4537	&	    1.91$^{+0.17}_{-0.11}$  	&	1.76 	&	1892	&	  3.04$\%$  \\
Z 535-12  	&	0.048 	&	7.36 	&	44.75 	&	$zr$	&	6080	&	    2.00$^{+0.18}_{-0.12}$  	&	1.76 	&	1887	&	  3.05$\%$  \\
NGC 5548*  	&	0.017 	&	7.82 	&	44.48 	&	$R$	&	6395	&	    2.40$^{+0.20}_{-0.20}$  	&	2.34 	&	4677	&	  4.68$\%$  \\
DES J021822.51-043036.0*  	&	0.823 	&	6.60 	&	44.49 	&	$g$	&	2579	&	    1.90$^{+0.30}_{-0.30}$  	&	1.08 	&	1212	&	  0.99$\%$  \\
SDSS J025007.03+002525.3*  	&	0.198 	&	7.96 	&	45.17 	&	$g$	&	3926	&	    2.20$^{+0.20}_{-0.20}$  	&	2.20 	&	6232	&	  2.54$\%$  \\
SDSS J153425.58+040806.7*  	&	0.040 	&	5.10 	&	42.46 	&	$r$	&	5940	&	    1.50$^{+0.50}_{-0.50}$  	&	0.39 	&	508	&	  0.48$\%$  \\
SDSS J160531.85+174826.3*  	&	0.032 	&	5.20 	&	42.14 	&	$r$	&	5985	&	    0.90$^{+0.20}_{-0.20}$  	&	0.63 	&	533	&	  0.80$\%$  \\
NGC 4151*  	&	0.003 	&	7.71 	&	44.06 	&	$g$	&	4754	&	    2.73$^{+0.26}_{-0.10}$  	&	2.41 	&	8118	&	  3.17$\%$  \\
NGC 7469*  	&	0.016 	&	7.04 	&	44.68 	&	$R$	&	5100	&	    2.40$^{+0.22}_{-0.16}$  	&	1.46 	&	6868	&	  0.42$\%$  \\
NGC 3516 high*  	&	0.009 	&	7.67 	&	44.08 	&	5100	&	5100	&	    2.62$^{+0.42}_{-0.26}$  	&	2.36 	&	4112	&	  5.57$\%$  \\
Mrk 335*  	&	0.026 	&	7.23 	&	44.24 	&	 $UVW$2	&	2066	&	    1.95$^{+0.09}_{-0.07}$  	&	1.84 	&	4871	&	  1.42$\%$  \\
Zw 229-15*  	&	0.028 	&	6.91 	&	44.10 	&	$Kepler$	&	6423	&	    1.96$^{+0.01}_{-0.01}$  	&	1.57 	&	1174	&	  3.16$\%$  \\
WPV S007*  	&	0.029 	&	6.60 	&	44.29 	&	$UVM$2	&	2527	&	    2.31$^{+0.13}_{-0.18}$  	&	1.17 	&	4397	&	  0.34$\%$  \\

\end{longtable*}
\noindent \textbf{Column notes}: Col. (1): Source name (sources with * represent objects that are not selected from the \texttt{Swift} BAT sample); Col. (2): The redshift; Col. (3): The black hole mass; Col. (4): The bolometric luminosity; Col. (5): Band Col. (6): The rest-frame wavelength; Col. (7): The rest-frame best-fitting damping timescale; Col. (8): The theoretical damping timescale; Col. (9): The baseline in the rest-frame; Col. (10): The $\tau_{\mathrm{th,Kelly+2009}}$-to-baseline ratio. The redshift, black hole mass, and the bolometric luminosity are adopt from BASS DR2 \citep[][]{Koss2022a,Koss2022b}. The samples with the same source name and the different $\lambda_{\mathrm{rest}}$ correspond to the different band light curves of the same source. In comparison with Table \ref{Tab2}, $\log_{10}(\tau_{\mathrm{obs}})$ for the same light curve may be slightly different because $\log_{10}(\tau_{\mathrm{obs}})$ in Tables \ref{Tab1} and \ref{Tab2} were obtained from the MCMC code. Similarly, the baseline for the same light curve may be slightly different due to the slightly different outlier rejections in each DRW MCMC fit.

\setlength\tabcolsep{1pt}
\begin{longtable*}{cccccccccc}
\caption{Sample with $\tau_{\mathrm{th,Zhou+2024}}/\mathrm{baseline} < 10\% $}

\label{Tab2}\\

\hline\hline
Object & $z$ & $\log_{10}(M_{\mathrm{BH}})$ & $\log_{10}(L_{\mathrm{bol}})$ & Band & $\lambda_{\mathrm{rest}}$ & $\log_{10}(\tau_{\mathrm{obs}})$ & $\log_{10}(\tau_{\mathrm{th,Zhou+2024}})$ & baseline & $\frac{\tau_{\mathrm{th,Zhou+2024}}}{\mathrm{baseline}}$  \\
& & $[M_{\odot}]$ & $[\mathrm{erg\, s^{-1} }]$ & & $[\mathrm{\AA}]$ & [days] & [days] & [days] & \\
\hline
\endhead
\hline\hline
\multicolumn{10}{r}{\textit{Continued on next page}} \\ 
\endfoot
\hline\hline
\endlastfoot
2MASS J06543417+0703210	&	0.024 	&	6.74 	&	44.23 	&	$zg$	&	4642	&	2.51$^{+0.43}_{-0.25}$&	2.24 	&	1806	&	9.62$\% $	\\
Arp 151	&	0.021 	&	6.67 	&	44.11 	&	$zg$	&	4657	&	1.99$^{+0.19}_{-0.13}$&	2.16 	&	1859	&	7.78$\% $	\\
CGM W5-04382	&	0.019 	&	7.43 	&	43.36 	&	$zg$	&	4663	&	2.24$^{+0.32}_{-0.20}$&	1.67 	&	1886	&	2.48$\% $	\\
CGM W5-04382	&	0.019 	&	7.43 	&	43.36 	&	$zr$	&	6249	&	0.93$^{+0.08}_{-0.08}$&	1.82 	&	1893	&	3.49$\% $	\\
CTS 103	&	0.012 	&	6.11 	&	43.63 	&	$zg$	&	4696	&	1.80$^{+0.25}_{-0.23}$&	1.85 	&	1893	&	3.74$\% $	\\
CTS 103	&	0.012 	&	6.11 	&	43.63 	&	$zr$	&	6293	&	2.12$^{+0.33}_{-0.21}$&	2.00 	&	1904	&	5.25$\% $	\\
ESO 424-12	&	0.010 	&	6.07 	&	43.48 	&	$zg$	&	4707	&	0.84$^{+0.11}_{-0.11}$&	1.75 	&	1798	&	3.13$\% $	\\
ESO 509-38	&	0.026 	&	8.32 	&	44.14 	&	$zg$	&	4632	&	2.06$^{+0.27}_{-0.20}$&	2.18 	&	1863	&	8.12$\% $	\\
ESO 548-81	&	0.014 	&	7.96 	&	44.16 	&	$zg$	&	4686	&	1.96$^{+0.31}_{-0.20}$&	2.19 	&	1583	&	9.78$\% $	\\
LED A75258	&	0.016 	&	7.02 	&	43.74 	&	$zg$	&	4679	&	1.80$^{+0.16}_{-0.12}$&	1.92 	&	1798	&	4.62$\% $	\\
LED A75258	&	0.016 	&	7.02 	&	43.74 	&	$zr$	&	6269	&	1.96$^{+0.22}_{-0.15}$&	2.07 	&	1656	&	7.09$\% $	\\
LED A90334	&	0.010 	&	6.61 	&	43.61 	&	$zg$	&	4704	&	1.30$^{+0.10}_{-0.09}$&	1.84 	&	1902	&	3.64$\% $	\\
LED A90334	&	0.010 	&	6.61 	&	43.61 	&	$zr$	&	6304	&	1.39$^{+0.10}_{-0.08}$&	1.99 	&	1890	&	5.17$\% $	\\
MCG+4-6-43	&	0.033 	&	7.69 	&	44.19 	&	$zg$	&	4599	&	2.27$^{+0.41}_{-0.21}$&	2.21 	&	1635	&	9.92$\% $	\\
Mrk 1044	&	0.017 	&	6.45 	&	43.72 	&	$zg$	&	4672	&	1.96$^{+0.25}_{-0.17}$&	1.91 	&	1632	&	4.98$\% $	\\
Mrk 1044	&	0.017 	&	6.45 	&	43.72 	&	$zr$	&	6261	&	1.82$^{+0.21}_{-0.14}$&	2.06 	&	1629	&	7.05$\% $	\\
Mrk 1310	&	0.019 	&	6.21 	&	43.82 	&	$zg$	&	4662	&	1.82$^{+0.21}_{-0.16}$&	1.97 	&	1865	&	5.01$\% $	\\
Mrk 352	&	0.015 	&	7.56 	&	43.96 	&	$zg$	&	4684	&	1.91$^{+0.19}_{-0.14}$&	2.06 	&	1687	&	6.81$\% $	\\
Mrk 352	&	0.015 	&	7.56 	&	43.96 	&	$zr$	&	6277	&	1.91$^{+0.17}_{-0.13}$&	2.21 	&	1825	&	8.89$\% $	\\
Mrk 359	&	0.017 	&	6.05 	&	43.75 	&	$zg$	&	4675	&	1.09$^{+0.09}_{-0.08}$&	1.93 	&	1672	&	5.09$\% $	\\
Mrk 359	&	0.017 	&	6.05 	&	43.75 	&	$zr$	&	6264	&	0.92$^{+0.10}_{-0.09}$&	2.08 	&	1660	&	7.24$\% $	\\
Mrk 595	&	0.027 	&	7.41 	&	44.10 	&	$zg$	&	4627	&	2.14$^{+0.33}_{-0.19}$&	2.15 	&	1619	&	8.72$\% $	\\
NGC 2885	&	0.026 	&	7.91 	&	44.20 	&	$zg$	&	4631	&	1.36$^{+0.10}_{-0.09}$&	2.22 	&	1831	&	9.07$\% $	\\
NGC 3080	&	0.035 	&	6.84 	&	44.01 	&	$zg$	&	4590	&	2.08$^{+0.25}_{-0.16}$&	2.10 	&	1832	&	6.87$\% $	\\
NGC 3080	&	0.035 	&	6.84 	&	44.01 	&	$zr$	&	6151	&	2.00$^{+0.22}_{-0.15}$&	2.25 	&	1830	&	9.72$\% $	\\
NGC 3227	&	0.003 	&	6.77 	&	43.67 	&	$zg$	&	4737	&	1.80$^{+0.18}_{-0.14}$&	1.87 	&	1908	&	3.89$\% $	\\
NGC 3227	&	0.003 	&	6.77 	&	43.67 	&	$zr$	&	6348	&	1.84$^{+0.17}_{-0.13}$&	2.03 	&	1893	&	5.66$\% $	\\
NGC 3822	&	0.019 	&	7.43 	&	43.91 	&	$zg$	&	4662	&	1.61$^{+0.18}_{-0.15}$&	2.03 	&	1876	&	5.71$\% $	\\
NGC 3822	&	0.019 	&	7.43 	&	43.91 	&	$zr$	&	6247	&	1.52$^{+0.15}_{-0.13}$&	2.18 	&	1864	&	8.12$\% $	\\
NGC 4051	&	0.002 	&	6.13 	&	42.59 	&	$zg$	&	4743	&	1.51$^{+0.11}_{-0.09}$&	1.17 	&	1920	&	0.77$\% $	\\
NGC 4051	&	0.002 	&	6.13 	&	42.59 	&	$zr$	&	6356	&	1.65$^{+0.13}_{-0.11}$&	1.32 	&	1883	&	1.11$\% $	\\
NGC 4235	&	0.008 	&	7.28 	&	43.26 	&	$zr$	&	6319	&	1.89$^{+0.21}_{-0.16}$&	1.76 	&	1811	&	3.18$\% $	\\
NGC 4253	&	0.013 	&	6.82 	&	43.75 	&	$zg$	&	4692	&	2.06$^{+0.22}_{-0.14}$&	1.93 	&	1901	&	4.48$\% $	\\
NGC 4253	&	0.013 	&	6.82 	&	43.75 	&	$zr$	&	6288	&	2.16$^{+0.25}_{-0.15}$&	2.08 	&	1886	&	6.38$\% $	\\
NGC 4395	&	0.001 	&	5.45 	&	41.68 	&	$zg$	&	4748	&	0.95$^{+0.07}_{-0.07}$&	0.58 	&	1924	&	0.20$\% $	\\
NGC 4395	&	0.001 	&	5.45 	&	41.68 	&	$zr$	&	6362	&	0.99$^{+0.07}_{-0.07}$&	0.73 	&	1872	&	0.29$\% $	\\
NGC 4593	&	0.008 	&	6.88 	&	44.00 	&	$zg$	&	4714	&	2.38$^{+0.48}_{-0.29}$&	2.09 	&	1908	&	6.45$\% $	\\
NGC 4593	&	0.008 	&	6.88 	&	44.00 	&	$zr$	&	6316	&	1.64$^{+0.18}_{-0.15}$&	2.24 	&	1892	&	9.18$\% $	\\
NGC 4748	&	0.014 	&	6.41 	&	43.63 	&	$zg$	&	4687	&	2.08$^{+0.27}_{-0.19}$&	1.85 	&	1885	&	3.76$\% $	\\
NGC 4748	&	0.014 	&	6.41 	&	43.63 	&	$zr$	&	6280	&	2.28$^{+0.38}_{-0.22}$&	2.00 	&	1882	&	5.31$\% $	\\
NGC 6814	&	0.006 	&	7.04 	&	43.52 	&	$zr$	&	6332	&	1.11$^{+0.18}_{-0.16}$&	1.93 	&	932	&	9.13$\% $	\\
NGC 7214	&	0.023 	&	7.44 	&	43.99 	&	$zg$	&	4648	&	1.85$^{+0.21}_{-0.16}$&	2.08 	&	1808	&	6.65$\% $	\\
NGC 7811	&	0.025 	&	6.70 	&	44.01 	&	$zg$	&	4635	&	1.87$^{+0.20}_{-0.14}$&	2.10 	&	1806	&	6.97$\% $	\\
UGC 12138	&	0.025 	&	7.08 	&	44.21 	&	$zg$	&	4637	&	2.49$^{+0.47}_{-0.24}$&	2.23 	&	1828	&	9.29$\% $	\\
UGC 3478	&	0.013 	&	6.06 	&	43.40 	&	$zg$	&	4693	&	1.99$^{+0.25}_{-0.18}$&	1.70 	&	1821	&	2.75$\% $	\\
UGC 3601	&	0.017 	&	7.33 	&	43.99 	&	$zg$	&	4673	&	1.50$^{+0.12}_{-0.10}$&	2.08 	&	1830	&	6.57$\% $	\\
UGC 3601	&	0.017 	&	7.33 	&	43.99 	&	$zr$	&	6261	&	1.71$^{+0.15}_{-0.11}$&	2.23 	&	1820	&	9.33$\% $	\\
UGC 6728	&	0.006 	&	5.55 	&	43.16 	&	$zg$	&	4725	&	1.11$^{+0.08}_{-0.07}$&	1.54 	&	1890	&	1.83$\% $	\\
UGC 6728	&	0.006 	&	5.55 	&	43.16 	&	$zr$	&	6331	&	1.16$^{+0.09}_{-0.08}$&	1.69 	&	1895	&	2.59$\% $	\\
Z 122-55	&	0.021 	&	7.13 	&	43.93 	&	$zg$	&	4654	&	2.04$^{+0.25}_{-0.16}$&	2.04 	&	1850	&	5.93$\% $	\\
Z 122-55	&	0.021 	&	7.13 	&	43.93 	&	$zr$	&	6236	&	1.85$^{+0.17}_{-0.13}$&	2.19 	&	1855	&	8.35$\% $	\\
Z 215-46	&	0.023 	&	7.50 	&	44.22 	&	$zg$	&	4646	&	2.64$^{+0.54}_{-0.29}$&	2.23 	&	1879	&	9.04$\% $	\\
Z 493-2	&	0.025 	&	7.44 	&	44.20 	&	$zg$	&	4638	&	1.83$^{+0.14}_{-0.11}$&	2.22 	&	1851	&	8.96$\% $	\\
NGC 5548*	&	0.017 	&	7.82 	&	44.48 	&	$R$	&	6395	&	2.40$^{+0.20}_{-0.20}$&	2.55 	&	4676	&	7.59$\% $	\\
DES J021822.51-043036.0*	&	0.823 	&	6.60 	&	44.49 	&	$g$	&	2579	&	1.90$^{+0.30}_{-0.30}$&	2.09 	&	1211	&	10.16$\% $	\\
SDSS J025007.03+002525.3*	&	0.198 	&	7.96 	&	45.17 	&	$g$	&	3926	&	2.20$^{+0.20}_{-0.20}$&	2.75 	&	6231	&	9.02$\% $	\\
SDSS J153425.58+040806.7*	&	0.040 	&	5.10 	&	42.46 	&	$r$	&	5940	&	1.50$^{+0.50}_{-0.50}$&	1.20 	&	508	&	3.12$\% $	\\
SDSS J160531.85+174826.3*	&	0.032 	&	5.20 	&	42.14 	&	$r$	&	5985	&	0.90$^{+0.20}_{-0.20}$&	1.00 	&	532	&	1.88$\% $	\\
NGC 4151*	&	0.003 	&	7.71 	&	44.06 	&	$g$	&	4754	&	2.73$^{+0.26}_{-0.10}$&	2.12 	&	8118	&	1.62$\% $	\\
NGC 7469*	&	0.016 	&	7.04 	&	44.68 	&	$R$	&	5100	&	2.40$^{+0.22}_{-0.16}$&	2.56 	&	6868	&	5.29$\% $	\\
NGC 3516 high*	&	0.009 	&	7.67 	&	44.08 	&	5100	&	5100	&	2.62$^{+0.42}_{-0.26}$&	2.17 	&	4112	&	3.60$\% $	\\
NGC 3516 low*	&	0.009 	&	7.67 	&	43.43 	&	5100	&	4728	&	1.75$^{+0.16}_{-0.12}$&	1.71 	&	1085	&	4.73$\% $	\\
Mrk 335*	&	0.026 	&	7.23 	&	44.24 	&	$UVW$2	&	2066	&	1.96$^{+0.09}_{-0.07}$&	1.81 	&	4871	&	1.33$\% $	\\
WPV S007*	&	0.029 	&	6.60 	&	44.29 	&	$UVM$2	&	2527	&	2.32$^{+0.21}_{-0.13}$&	1.80 	&	4396	&	1.44$\% $	\\
NGC 5273 g high	&	0.004 	&	6.66 	&	41.97 	&	$zg$	&	4734	&	1.23$^{+0.25}_{-0.29}$&	0.77 	&	501	&	1.17$\% $	\\
NGC 5273 g low	&	0.004 	&	6.66 	&	41.77 	&	$zg$	&	4734	&	1.29$^{+0.15}_{-0.12}$&	0.64 	&	1222	&	0.36$\% $	\\
NGC 5273 r high	&	0.004 	&	6.66 	&	42.07 	&	$zr$	&	6344	&	0.50$^{+0.30}_{-0.25}$&	0.98 	&	474	&	2.01$\% $	\\
NGC 5273 r low	&	0.004 	&	6.66 	&	41.91 	&	$zr$	&	6344	&	1.53$^{+0.16}_{-0.13}$&	0.88 	&	1258	&	0.60$\% $	\\

\end{longtable*}
\noindent \textbf{Column notes}: Col. (1): Source name (sources with * represent objects that are not selected from the \texttt{Swift} BAT sample); Col. (2): The redshift; Col. (3): The black hole mass; Col. (4): The bolometric luminosity; Col. (5): Band Col. (6): The rest-frame wavelength; Col. (7): The rest-frame best-fitting damping timescale; Col. (8): The theoretical damping timescale; Col. (9): The baseline in the rest-frame; Col. (10): The $\tau_{\mathrm{th,Kelly+2009}}$-to-baseline ratio. The redshift, black hole mass, and the bolometric luminosity are adopted from BASS DR2 \citep[][]{Koss2022a,Koss2022b}. The samples with the same source name and the different $\lambda_{\mathrm{rest}}$ correspond to the different band light curves of the same source. In comparison with Table \ref{Tab1}, $\log_{10}(\tau_{\mathrm{obs}})$ for the same light curve may be slightly different because $\log_{10}(\tau_{\mathrm{obs}})$ in Tables \ref{Tab1} and \ref{Tab2} were obtained from the MCMC code. Similarly, the baseline for the same light curve may be slightly different due to the slightly different outlier rejections in each DRW MCMC fit.

\section{Results of DRW fitting to the $zr$ band light curve of ESO 424-12}

\begin{figure}[H]
\plotone{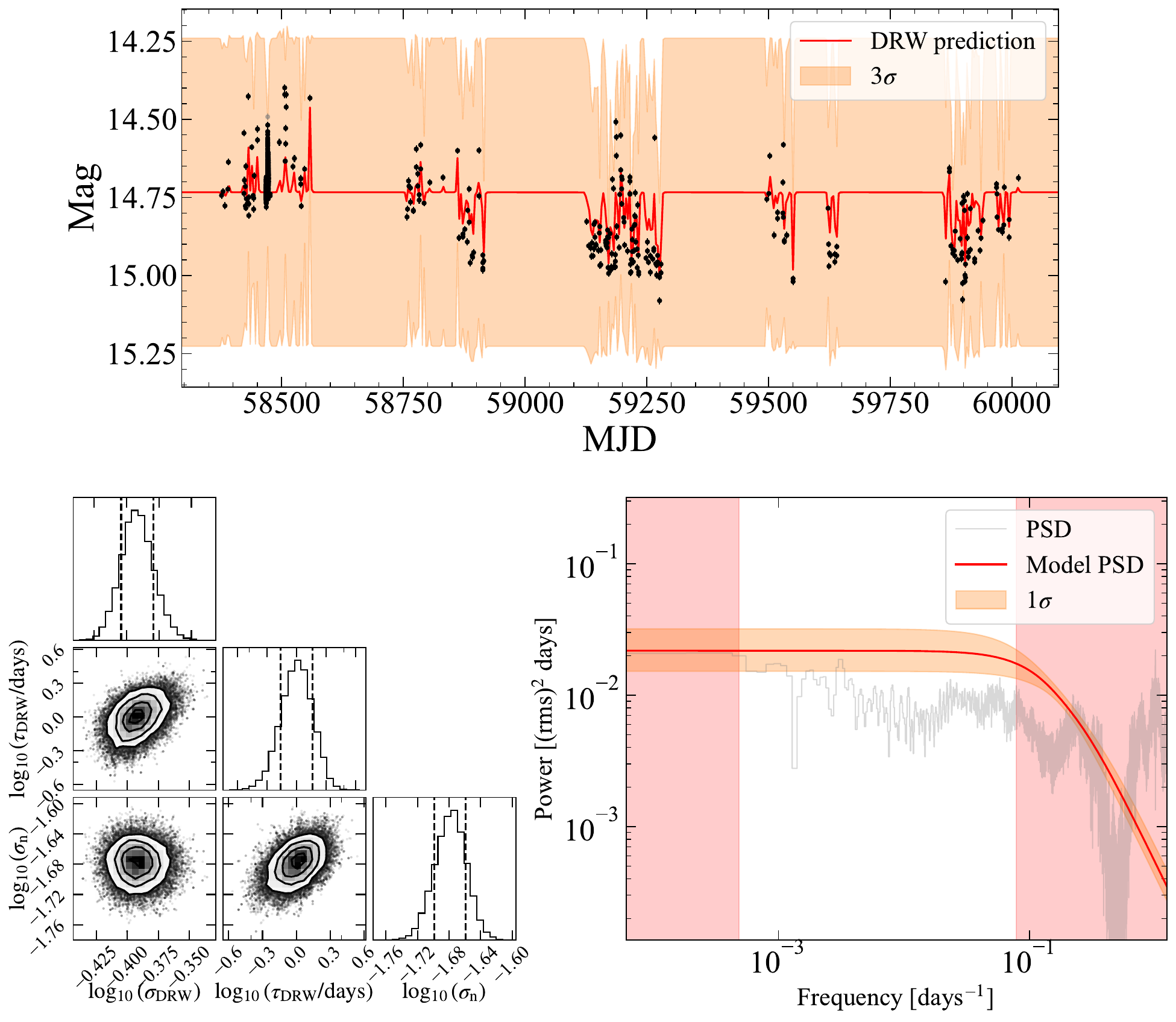}
\caption{Fitting the $zr$ band light curve (in the observed-frame) of ESO 424-12 with the DRW model via the \textbf{\texttt{taufit}} code. The meaning of the data points and lines in this
figure are the same as in Fig.~\ref{Fig1}. There are unusually dense observations (about MJD = 58500) for ESO 424-12, which leads to the damping timescale of this light curve being unusually small, i.e., about 1 day.
\label{FigB1}}
\end{figure}


\bibliography{ref}{}
\bibliographystyle{aasjournal}



\end{document}